\pdfoutput=0
\documentclass[11pt]{article}
\usepackage{axodraw}
\usepackage{epsfig}
\usepackage{amsfonts}
\usepackage{amsmath}
\usepackage{bbm,bm}
\usepackage{cite}
\allowdisplaybreaks[1]

\input pix.sty

\def\undertilde#1{\mathop{\vtop{\ialign{##\cr$\textstyle{#1}$\cr%
\noalign{\kern1pt\nointerlineskip}\hfil$\mathchar"0365$\hfil\cr}}}}
\def\wideundertilde#1{\mathop{\vtop{\ialign{##\cr$\textstyle{#1}$\cr%
\noalign{\kern1pt\nointerlineskip}\hfil$\mathchar"0367$\hfil\cr}}}}

\newcommand{\rate}{sph} 
\newcommand{\X}{x} 
\renewcommand{\vec}[1]{{\bf #1}}

\newcommand{\tinymsbar}{{\overline{\mbox{\tiny\rm{MS}}}}}
\newcommand{\Lambdamsbar}{{\Lambda_\tinymsbar}}
\newcommand{\alphas}{\alpha} 
\newcommand{\Nf}{N_{\rm f}}
\newcommand{\Nc}{N_{\rm c}}

\renewcommand{\S}{\rmii{S}}

\newcommand{\dA}{d_\rmii{A}}

\newcommand{\mD}{m_\rmii{D}}

\newcommand{\rmO}{{\mathcal{O}}}

\def\lsi{\raise0.3ex\hbox{$<$\kern-0.75em\raise-1.1ex\hbox{$\sim$}}}
\def\gsi{\raise0.3ex\hbox{$>$\kern-0.75em\raise-1.1ex\hbox{$\sim$}}}
\newcommand{\lsim}{\mathop{\lsi}}
\newcommand{\gsim}{\mathop{\gsi}}

\newcommand{\nB}{n_\rmii{B}}
\newcommand{\rmii}[1]{{\mbox{\tiny\rm{#1}}}}
\newcommand{\rmiii}[1]{{\mbox{\tiny{$\scriptstyle{\rm#1}$}}}}

\newcommand{\im}{\mathop{\mbox{Im}}}
\newcommand{\Tint}[1]{{\hbox{$\sum$}\!\!\!\!\!\!\!\int\,}_{\!\!\!\!\raise-0.9ex\hbox{$\scriptstyle{#1}$}}}
\newcommand{\Tinti}[1]{{{\Sigma}\!\!\!\!\raise0.3ex\hbox{$\int$}_\rmii{${#1}$}}}
\newcommand{\Tintip}[1]{{{\Sigma'}\!\!\!\!\!\raise0.3ex\hbox{$\int$}_\rmii{${#1}$}}}
\newcommand{\bi}{\begin{itemize}}
\newcommand{\ei}{\end{itemize}}
\newcommand{\hide}[1]{ }

\def\ring{\mathaccent"7017} 

\makeatletter \@addtoreset{equation}{section} \makeatother
\renewcommand{\theequation}{\arabic{section}.\arabic{equation}}
\makeatletter
\renewcommand\section{\@startsection {section}{1}{\z@}%
                                   {-5.5ex \@plus -1ex \@minus -.2ex}
                                   {2.3ex \@plus.2ex}%
                                   {\normalfont\large\bfseries}}
\renewcommand\subsection{\@startsection{subsection}{2}{\z@}%
                                     {-3.25ex\@plus -1ex \@minus -.2ex}%
                                     {1.5ex \@plus .2ex}%
                                     {\normalfont\normalsize\bfseries}}
\renewcommand\thesection {\@arabic\c@section}
\renewcommand\thesubsection   {\thesection.\@arabic\c@subsection}
\renewcommand{\@seccntformat}[1]{%
\csname the#1\endcsname.\hspace{1.0em}}
\makeatother

\begin{document}

\flushbottom


\begin{titlepage}

\begin{flushright}
November 2022
\end{flushright}
\begin{centering}
\vfill

{\Large{\bf
 Shape of the hot topological charge density spectral function 
}} 

\vspace{0.8cm}

M.~Laine\hspace*{0.2mm}$^\rmi{a}$, 
L.~Niemi$^\rmi{b,c}$, 
S.~Procacci\hspace*{0.2mm}$^\rmi{a}$, 
K.~Rummukainen$^\rmi{c}$
 
\vspace{0.8cm}

$^\rmi{a}$%
{\em
AEC, 
Institute for Theoretical Physics, 
University of Bern, \\ 
Sidlerstrasse 5, CH-3012 Bern, Switzerland \\}

\vspace{0.3cm}

$^\rmi{b}$%
{\em
Tsung-Dao Lee Institute, 
Shanghai Jiao Tong University, \\
Shanghai 200240, China \\}

\vspace{0.3cm}

$^\rmi{c}$%
{\em
Department of Physics and Helsinki Institute of Physics,\\ 
P.O.~Box 64, FI-00014 University of Helsinki, Finland \\}

\vspace*{0.8cm}

\mbox{\bf Abstract}
 
\end{centering}

\vspace*{0.3cm}
 
\noindent
After motivating an interest in the shape of the topological charge
density spectral function in hot Yang-Mills theories, we estimate it
with the help of thermally averaged classical real-time
simulations, for $\Nc^{ } = 2,3$. 
After subtracting a perturbative contribution at large
frequencies, we observe a non-trivial shape at small frequencies
(a dip rather than a peak),
interpolating smoothly towards the sphaleron rate at zero frequency. 
Possible frequency scales making an appearance
in this shape are discussed.  
Implications for warm axion inflation and reheating, 
and for imaginary-time lattice measurements 
of the strong sphaleron rate, are recapitulated.

\vfill

 
\end{titlepage}

\tableofcontents

%
\section{Introduction}
\la{se:intro}

The operator known as 
the topological charge density $\chi$
(cf.\ \eq\nr{chi})
plays a remarkable role in non-Abelian
quantum field theory. 
For the weak gauge group, it is sensitive to processes responsible for 
anomalous baryon plus lepton number violation~\cite{anomaly}. 
The violation rate is believed to be fast at high temperatures
and therefore important for baryogenesis~\cite{krs}, a fact that
has led to its detailed numerical determination~\cite{sphaleron}.
On the semiclassical level,  
the gauge field configurations mediating baryon plus lepton
number violation in the Higgs phase
are known as sphalerons~\cite{kli}.
Consequently the thermal rate is generically 
referred to as the sphaleron rate, 
or the Chern-Simons diffusion rate.   
 
An analogue of the sphaleron rate originating from the strong
gauge group may be of interest as well. It affects the
evolution of QCD axions~\cite{mms}, 
among the most studied dark matter candidates. 
It leads to a fast violation of chirality, 
which has motivated investigations in the context
of heavy ion collision experiments
(cf., e.g.,\ ref.~\cite{eucl} for recent work and references). 
Finally, the topological charge density plays an essential
role in models of natural inflation~\cite{ai,ai_em,ai_rev}.
Indeed numerical results for the strong sphaleron rate~\cite{mt}
have found use in that 
context~\cite{warm1,warm3,warm4,warm,warm45,warm5,warm6}, 
and may induce an efficient 
reheating mechanism as well~\cite{gravity}.

On the technical level, thermal rates are often referred to as
transport coefficients. Through so-called Kubo relations, they can be 
extracted from equilibrium 2-point correlation functions. Specifically, 
if 
$
 C^{ }_{\S}(\omega) 
$
denotes a Fourier transform of a 2-point correlator
$
 C^{ }_{\S}(t) 
$
with a specific time ordering 
(cf.\ \eq\nr{Delta}), 
then the sphaleron rate is given by 
$
 \Gamma^{ }_\rmi{\rate} = 
 \lim^{ }_{\,\omega\to 0}  C^{ }_{\S}(\omega) 
$ 
(for a discussion that does not rely on perturbative 
arguments or physical intuition, see ref.~\cite{son}). 

The focus of the current investigation is  
the {\em shape} of 
$
 C^{ }_{\S}(\omega) 
$, 
which contains additional information. 
The shape is relevant, for example, in theories in which
a scalar field $\varphi$, notably an inflaton as mentioned above, 
couples to a plasma through the operator 
$\mathcal{L}\supset - \varphi\chi / f^{ }_a$. Then it is 
the $\omega$-dependence
that determines the efficiency of the friction that the 
plasma exerts on $\varphi$
in different temperature domains~\cite{warm}. 
Another example is that estimating the rate of anomalous
chirality violation in QCD requires imaginary-time lattice
simulations~\cite{eucl}, and then it would be important to 
have an ansatz for the $\omega$-dependence.  
Even though the QCD coupling is so large that our methods
are not reliable on a quantitative level, it is  
believed that large couplings smoothen spectral functions
rather than insert sharp features in them, whereby our
results could still be helpful on the qualitative level.

The plan of this paper is the following. 
After formulating the technical problem 
(\se\ref{se:formulation}), 
we review the framework of thermally averaged
classical simulations that we use for addressing it (\se\ref{se:clgt}), 
and describe practical details of the
numerical effort (\se\ref{se:simu}).
Subsequently we discuss our 
slightly unexpected
results for the sphaleron rate (\se\ref{se:cs}). 
The main part is the analysis of the shape of the spectral
function (\se\ref{se:shape}), 
after which conclusions and an outlook can be offered
(\se\ref{se:concl}). 
Details of a perturbative computation 
are deferred to appendix~A.

%
\section{Formulation of the problem}
\la{se:formulation} 

In pure SU($\Nc^{ }$) Yang-Mills theory, 
the topological charge density is defined as 
\be
 \chi \;\equiv\;
 c^{ }_{\chi} \, \epsilon^{\mu\nu\rho\sigma}_{ }
 g^2 F^{a}_{\mu\nu} F^{a}_{\rho\sigma}
 \;, 
 \quad
 a \in \{ 1, ..., \Nc^2 - 1  \} 
 \;, \quad
 c^{ }_\chi \;\equiv\; \frac{1}{64\pi^2}
 \;, \la{chi}
\ee
where 
$T^a_{ }F^a_{\mu\nu} = [D^{ }_\mu,D^{ }_\nu]/(ig)$ 
is the Yang-Mills field strength; 
$
 D^{ }_\mu \equiv \partial^{ }_\mu + i g T^a_{ }A^a_\mu
$
is a covariant derivative; 
$g^2 \equiv 4\pi \alphas$ is the Yang-Mills coupling;
and $T^a_{ }$ are Hermitean generators of SU($\Nc^{ }$), 
normalized as $\tr[T^a_{ }T^b_{ }] = \delta^{ab}_{ }/2$.
The topological charge density is a peculiar quantity, 
in that for smooth gauge configurations
it is a total derivative, evaluating to an integer after 
integration over spacetime. 
In quantum field theory, it is simply a local pseudoscalar operator, 
which displays non-trivial correlation functions at all time 
and distance scales. 

In this study we are interested in real-time correlation functions 
of the spatial average of~$\chi$, 
in thermal equilibrium at a temperature~$T$. 
Different time orderings yield different real-time correlation functions.
Theoretical discussions (and also the title of this paper)
often refer to a spectral function, which is defined as 
\be
 \rho(\omega) \; \equiv \; 
 \int_{-\infty}^{\infty} \! {\rm d}t\, e^{i\omega t} \, 
 \int_\vec{x} 
 \Bigl\langle
  \frac{1}{2} \bigl[ \chi(t,\vec{x}) \,,\, \chi(0,\vec{y}) \bigr]
 \Bigr\rangle
 \;. \la{rho}
\ee
The spectral function can alternatively be viewed 
as the imaginary part of a retarded correlator, 
$
 \rho(\omega) = \im C^{ }_\rmii{R}(\omega + i 0^+_{ })
$.
However, as discussed in \se\ref{se:intro}, 
physical observables are more directly related to a 
``statistical'', or time-symmetric 2-point correlator of $\chi$, 
\be
 C^{ }_{\S}(t) \; \equiv \; 
 \int_\vec{x} 
 \Bigl\langle
  \frac{1}{2} \bigl\{ \chi(t,\vec{x}) \,,\, \chi(0,\vec{y}) \bigr\}
 \Bigr\rangle
 \;, \quad
 C^{ }_{\S}(\omega) 
 \; \equiv \;
 \int_{-\infty}^{\infty} \! {\rm d}t\, e^{i\omega t} \, 
 C^{ }_{\S}(t)
 \;. \la{Delta}
\ee
Given that 
$
 C^{ }_{\S}(-t) = 
 C^{ }_{\S}(t)
$, 
the Fourier transform can equivalently be expressed as 
\be
 C^{ }_{\S}(\omega) 
 = 2 
 \int_{0}^{\infty} \! {\rm d}t\, \cos ({\omega t}) \, 
 C^{ }_{\S}(t)
 \;. \la{C_w}
\ee
A text-book proof, obtained by inserting complete sets of 
energy eigenstates in the thermal expectation values, shows that
the spectral and statistical correlators are related by 
\be
 C^{ }_{\S}(\omega)
 \; = \; 
 \bigl[ 1 + 2 \nB^{ }(\omega) \bigr]\, \rho(\omega)
 \;, \la{relation}
\ee
where 
$
 \nB^{ }(\omega) \equiv 1 / ( e^{\beta\omega} - 1 )
$ 
is the Bose distribution, with $\beta \equiv 1/T$. 
In the domain that we are interested
in, {\it viz.}\ $\omega \ll T$, the spectral function can thus
be obtained from the statistical correlator as 
\be
 \rho(\omega) \quad \stackrel{|\omega| \ll T}{ = } \quad
 \frac{\omega\,  C^{ }_{\S}(\omega) }{2T}
 \;. \la{relation2}
\ee

Now, the appearance of an anticommutator in \eq\nr{Delta} guarantees that 
$
  C^{ }_{\S}(\omega) 
$
has formally a classical limit, 
\be
  C^\rmi{(cl)}_{\S}(\omega)
 \; \equiv \;  
 \lim_{\hbar\to 0}
  C^{ }_{\S}(\omega) 
 \;. \la{C_cl_w}
\ee
It turns out that in an interacting theory, the classical limit is 
singular~\cite{clgt3}, 
as classical field theory is plagued by Rayleigh-Jeans type
of ultraviolet (UV) divergences. 
However, if we keep the UV cutoff finite, by
introducing a {\em spatial} 
lattice discretization, then the classical limit 
exists. It is believed that studies with such a framework 
can reveal the magnitude of the transport coefficient  
$ \lim_{\omega\to 0} C^{ }_{\S}(\omega) $
at weak coupling, $\alpha\Nc^{ } \ll 1$~\cite{clgt1,old1,clgt2}.
The premise of the present investigation is 
that it should also be possible to  
use classical lattice gauge theory to 
estimate the shape of~$ C^{ }_{\S}(\omega) $, 
as long as we are in the Bose-enhanced domain $|\omega| \ll T$.

The UV divergences mentioned above distort physics 
at the frequency scale of the cutoff, $\omega \sim 1/a$. 
In order to alleviate this, we compute 
$
 C^\rmi{(cl)}_{\S}(\omega)
$
in leading-order perturbation theory for $\omega \sim 1/a$, 
where $a$ is the lattice spacing. In this UV regime, classical
lattice gauge theory is weakly coupled. 
The perturbative result can be subtracted from the 
lattice measurement,\footnote{%
  The perturbative expression vanishes as $\omega\to 0$ and therefore
  plays no role in the estimation of the transport coefficient
  $
   \lim_{\omega\to 0} C^{ }_{\S}(\omega) 
  $. 
 } 
whose goal is to estimate
$
 C^\rmi{(cl)}_{\S}(\omega)
$
non-perturbatively at 
$ 
 \omega \sim  \{ \alpha^2\Nc^2 T^2 a , \alpha\Nc^{ } T \}
$. 
The first scale replaces the physical infrared 
(IR) scale $\sim \alphas^2\Nc^2 T$ in classical
lattice gauge theory~\cite{clgt4}. 
The second scale represents the colour-magnetic screening scale, 
but it can also affect real-time phenomena, given that space-like
separated real-time fluctuations 
can to a good approximation be treated as equal-time ones. 
With a suitable re-interpretation
of these IR frequency scales, the results then arguably apply, 
on a qualitative level, 
to the continuum problem as well~\cite{clgt4,mr}
(cf.\ \ses\ref{se:shape} and \ref{se:concl}).\footnote{%
 In recent numerical determinations of the sphaleron rate, 
 a different logic is followed
 (cf.,\ e.g.,\ refs.~\cite{sphaleron,mt} and references therein). 
 By integrating out the scale $\sim gT$, it is possible to derive a simplified 
 Langevin description for the IR dynamics~\cite{db1,db2}. 
 A great advantage of this setup is that it is UV-finite~\cite{ay1,m1}.
 However, there are drawbacks, 
 namely that 
 the simplest form of the theory, 
 used for numerical simulations~\cite{sphaleron,mt},
 applies only to the smallest frequencies,
 whereas we would like to 
 resolve the shape up to somewhat larger frequencies;
 and that it involves
 an expansion not only in $g$, but also in $1/\ln(1/g)$~\cite{ay2}.
 } 

%
\section{Definition of classical lattice gauge theory}
\la{se:clgt}

We consider a theory discretized in spatial directions and with 
a continuous time coordinate~\cite{kogut,ml}. Quantizing this theory
in the gauge $A^{a}_0 = 0$ and then taking the classical limit,  
yields the partition function
(cf.,\ e.g.,\ ref.~\cite{mr} and references therein) 
\be
 Z^\rmi{(cl)}_{ } = 
 \int\! 
 \mathcal{D} U^{ }_i\, \mathcal{D}\mathcal{E}^{ }_i\, \delta(G) 
 \exp \biggl\{
    - \frac{1}{g^2 T a} \sum_\vec{x}
      \biggl[
      \sum_{i,j} \tr \bigl( \mathbbm{1} - P^{ }_{ij} \bigr) 
      + 
      \sum_i \tr\bigl( \mathcal{E}_i^2 \bigr)
      \biggr] 
 \biggr\} 
 \;, \la{Z_cl}
\ee
where $U^{ }_i$ are link matrices;
$\mathcal{E}^{ }_i$ the corresponding canonical momenta; \linebreak
$
   P^{ }_{ij}(\vec{x}) = 
   U^{ }_{i}(\vec{x}) U^{ }_{j}(\vec{x} + a \vec{i})
   U^{\dagger}_{i}(\vec{x} + a \vec{j})U^\dagger_{j}(\vec{x}) 
$ 
is a plaquette;
\be
 G(\vec{x}) \;\equiv\; \sum_i \bigl[ 
 \mathcal{E}^{ }_i(\vec{x}) - 
 U^\dagger_i(\vec{x} - a \vec{i})
 \mathcal{E}^{ }_i(\vec{x} - a \vec{i})
 U^{ }_i(\vec{x} - a \vec{i}) \bigr]
 \la{gauss}
\ee
are Gauss law operators, set to zero at every location
$\vec{x}$ by the constraints in \eq\nr{Z_cl};
and $\vec{i}$ is a unit vector in the $i$-direction. 
The equations of motion read
\ba
 a\, \partial^{ }_t U^{ }_i(\X) & = &  i \mathcal{E}^{ }_i(\X) U^{ }_i(\X)
 \;, \la{eom1} \\[2mm]
 a\, \partial^{ }_t \mathcal{E}^b_i(\X) & = & 
 2 \sum_{j\neq i} \im \tr \bigl\{ T^b_{ }
  \bigl[\,
   P^{ }_{ji}(\X) + P^{ }_{-ji}(\X)  
  \,\bigr]\bigr\} 
 \;, \la{eom2}
\ea
where $U^{ }_{-j}(\X) \equiv U^\dagger_{j}(\X - a \vec{j})$, 
and $\X \equiv (t,\vec{x})$.
The Gauss law at each position $\vec{x}$, 
and the Hamiltonian, are constants of motion. 
For later reference we also note that $\mathcal{E}^{ }_i$ is related 
to a continuum electric field $E^{ }_i$ as 
$
 \mathcal{E}^{ }_i = a^2 g E^{ }_i
$, 
implying that the Hamiltonian 
(appearing as $e^{-H/T}$)
contains
\be
 H \; \supset \; 
 \frac{1}{g^2 a} 
 \sum_\vec{x} \sum_i 
 \tr \bigl( \mathcal{E}_i^2 \bigr)
 \; = \; 
 \sum_\vec{x} a^3 \sum_i \tr\bigl( E_i^2 \bigr)
 \;. \la{Ei_1}
\ee

Now, the continuum operator from \eq\nr{chi} can be written as 
\be
 \chi
 \; = \;
 4 c^{ }_{\chi} \, \epsilon^{ }_{ijk}\,
 g^2 F^{b}_{0 i} F^{b}_{jk}
 \; = \;
 4 c^{ }_{\chi} \, \epsilon^{ }_{ijk}\,
 g^2 E^{b}_{i} F^{b}_{jk}
 \; = \;
 - 8 i c^{ }_{\chi} \, \epsilon^{ }_{ijk}\,
 \tr \bigl( 
 g E^{ }_{i}\, i g F^{ }_{jk} \bigr) 
 \;. \la{chi2}
\ee
The electric part could be expressed in terms of 
$\mathcal{E}^{ }_i$, but for practical measurements 
it turns out to be important to symmetrize the discretization
(cf., e.g., ref.~\cite{old}).
According to \eq\nr{eom1}, the electric fields $\mathcal{E}^{ }_i(x)$
affect the evolution of the links placed 
between $x$ and $x + a \mathbf{i}$. We can thus imagine that electric
fields ``live'' at $x + \frac{ a \mathbf{i}}{2}$.
For the position $x$ it is best to associate
an electric field covariantly averaged from 
$x + \frac{ a \mathbf{i}}{2}$ and 
$x - \frac{ a \mathbf{i}}{2}$, 
\ba
 \mathcal{\overline{E}}^{ }_i(x) & \equiv & 
 \frac{1}{2} 
 \Bigl[\,  
 \mathcal{E}^{ }_i(x) +
   U^\dagger_i(x-a\mathbf{i}) 
   \mathcal{E}^{ }_i(x - a \mathbf{i}) U^{ }_i(x-a\mathbf{i})
 \,\Bigr]
 \;. \la{E_impr}  
\ea
The parts appearing here are the same as needed in the Gauss
law, cf.\ \eq\nr{gauss}, now just averaged over, 
rather than subtracted from each other. 

As far as the magnetic field goes, it is 
discretized by making use of a ``clover'', 
\ba
 ig {F}^{ }_{jk}(\X) & \equiv & 
 \frac{  Q^{ }_{jk}(\X) - Q^{ }_{kj}(\X)  }{8a^2} 
 \;, \la{Fjk} \\[2mm] 
 Q^{ }_{jk}(\X) & \equiv & 
 P^{ }_{jk}(\X) + P^{ }_{k-j}(\X) + P^{ }_{-j-k}(\X) + P^{ }_{-k j}(\X)
 \;. 
\ea
Making use of translational invariance, 
the measurement of \eq\nr{Delta} therefore originates from 
\ba
 a^5 C^\rmi{(cl)}_{\S}(t) & \equiv & 
 \frac{1}{N_s^3}
 \sum_{\vec{x},\vec{y}} 
 \Bigl\langle
   a^4 \chi^\rmi{(cl)}_{ }(t,\vec{x}) \, a^4 \chi^\rmi{(cl)}_{ }(0,\vec{y}) 
 \Bigr\rangle
 \;, \la{a5Ccl} \\
 a^4 \chi^\rmi{(cl)}_{ }  & \equiv &
  - 2 i c^{ }_\chi \, 
 \epsilon^{ }_{ijk} \, \tr\bigl(\,
    \mathcal{\overline{E}}^{ }_i \,
    Q^{ }_{jk} \,
                          \bigr)
 \;. \la{chi_impr}
\ea
On the right-hand side of \eq\nr{a5Ccl}, 
factors of $a$ from the inverse volume
and two summation measures have been combined as 
$a^{-3}_{ }(a^3_{ })^2 a^5_{ } = (a^4_{ })^2_{ }$;
$N^{ }_s$ is the number of lattice points in spatial directions; 
the time dependence 
is obtained by solving the equations of motion in
\eqs\nr{eom1} and \nr{eom2}; and initial conditions are 
generated according to the weight in \eq\nr{Z_cl}. 

We note that, unlike in continuum, 
\eq\nr{chi_impr} is not ``topological'', 
i.e.\ a total derivative (or a difference). Much has been said about
this in the literature, but we are not worried, 
as the operator still has the correct IR properties 
at small energies and momenta, and computable UV properties 
at energy scales $\sim 1/a$. More comments on this are
offered in \se\ref{se:cs}.

Once we have measured 
$
  C^\rmi{(cl)}_{\S}(t)
$
from \eq\nr{a5Ccl} 
and taken a Fourier transform according to \eq\nr{C_w},  
the results are
conveniently normalized as~\cite{mr,mt}
\be
 \frac{ C^\rmi{(cl)}_{\S}(\omega) }{(\alphas T)^4_{ }}
 \; = \; 
 \frac{ a^4_{ } C^\rmi{(cl)}_{\S}(\omega)}
      { 16\, c_\chi^2\, (a g^2_{ }T)^4_{ } }
 \;. \la{normalization}
\ee

%
\section{Simulations and measurements}
\la{se:simu}

For practical measurements, 
the setup outlined above needs to 
be made concrete through a number of further ingredients: 
the time evolution needs to be discretized, a correctly
thermalized ensemble of initial configurations needs to be generated, 
observables need be averaged as much as possible to reduce noise, 
the spatial volume needs to be made large enough so 
that it has no practical effect,
and the equations of motion need to be solved for a long enough time 
that small frequencies can be addressed 
(including $\omega\to 0$ as needed for
$\Gamma^{ }_\rmi{\rate}$, cf.\ \eq\nr{Gamma_sph}).  
In this section we describe how these challenges can be overcome. 

\vspace*{0.3cm}

%
\subsubsection*{Time discretization}

For representing 
the time evolution in \eqs\nr{eom1} and \nr{eom2}, 
we choose a temporal lattice spacing $a^{ }_t \ll a$. 
In practice, after tests, we settled on 
$a^{ }_t = 0.02 a$, which turns out to yield sufficient
accuracy at a manageable cost.

Even if the links
and electric fields are placed 
at the same time in the operator
$\chi^\rmi{(cl)}_\rmi{ }$ 
in \eq\nr{chi_impr}, for a numerical evaluation
it is helpful to first evolve the links by $a^{ }_t/2$ in the
time direction. Subsequently, we leap-frog the evolutions of the links
and the electric fields, 
\ba
 U^{ }_i \biggl( x + \frac{a^{ }_t \mathbf{0}}{2}  \biggr)
 & \equiv & 
 \exp\biggl[ 
    i \frac{ a^{ }_t }{a}\, 
  \mathcal{\overline{E}}^{ }_i(x)
 \biggr]
 U^{ }_i \biggl( x - \frac{a^{ }_t \mathbf{0}}{2}  \biggr)
 \;, \la{dU_at} \\[2mm]
 \mathcal{E}^b_i\bigl(\X + a^{ }_t \mathbf{0} \bigr) & = & 
 \mathcal{E}^b_i\bigl(\X\bigr)  
 + \frac{ 2 a^{ }_t}{a}
  \sum_{j\neq i} \im \tr \biggl\{ T^b_{ }
  \biggl[\,
   P^{ }_{ji} \biggl( x + \frac{a^{ }_t \mathbf{0}}{2}  \biggr)
 + P^{ }_{-ji}\biggl( x + \frac{a^{ }_t \mathbf{0}}{2}  \biggr)  
  \,\biggr]\biggr\} 
 \;, \hspace*{5mm} \la{dE_at}
\ea
where $ \mathbf{0} $ is a unit vector in the time direction. 
For measuring $\chi^\rmi{(cl)}_\rmi{ }$, 
the Hamiltonian, 
or for checking the conservation
of the Gauss law constraints, the links and electric fields are brought
back to the same time, by performing an additional half-step
update of $U^{ }_i$.

It is important to stress that the Hamiltonian and the Gauss law
constraints are conserved for $a^{ }_t \ll a$
and in practice stay unchanged within our numerical resolution. Therefore, 
any configuration obtained after real-time evolution is an 
equally valid representative of the thermal ensemble. In other words,
the system is time-translation invariant. 

%
\subsubsection*{Thermalization}

The generation of a thermal ensemble representing \eq\nr{Z_cl} differs
from a standard lattice simulation by the need to satisfy the Gauss 
constraint at each position. To implement this, we have
followed the procedure described in ref.~\cite{gdm}, which is 
summarized here. One full thermalization ``sweep'' 
(over the spatial lattice) consists of the following steps: 

\begin{itemize}

\item[(i)]
We first pull $\{ \mathcal{E}^{ }_i(\vec{x}) \}$ from a Gaussian distribution
according to the last term in \eq\nr{Z_cl}, leaving the links
$\{ U^{ }_i (\vec{x}) \}$ untouched. 

\item[(ii)]
The Gauss law constraints $G(\vec{x})=0$ are subsequently
enforced by repeatedly modifying 
\be
 \mathcal{E}^{ }_i(\vec{x}) \to 
 \mathcal{E}^{ }_i(\vec{x}) - 
 \gamma \bigl[
          G(\vec{x})
   - U^{ }_i(\vec{x}) G(\vec{x} + a \vec{i} ) U^\dagger_i(\vec{x})
        \bigr]
 \;, \quad \gamma > 0
 \;, 
\ee
with the $G(\vec{x})$ recalculated at each step. This corresponds
to a gradient flow minimization of an auxiliary Hamiltonian 
$\sum_{\vec{x}} G^a_{ }(\vec{x}) G^a_{ }(\vec{x}) $~\cite{old1}. Following
ref.~\cite{gdm}, the update size $\gamma$ is alternated between
$5/48$ and $5/24$, and the process is repeated until the average
per site violation 
$\sum_{\vec{x}} G^a_{ }(\vec{x}) G^a_{ }(\vec{x})/N_s^3 $
is smaller than $10^{-12}_{ }$.

\item[(iii)]
Steps (i) and (ii) give a duly thermalized configuration
$\{ \mathcal{E}^{ }_i (\vec{x})\}$ for fixed $\{ U^{ }_i(\vec{x})\}$.
In order to generate new links, we allow energy to redistribute between
the $\{ \mathcal{E}^{ }_i (\vec{x})\}$ and $\{ U^{ }_i(\vec{x})\}$, 
by evolving the system with 
\eqs\nr{dU_at} and \nr{dE_at}, until time $t = a$. As mentioned above, 
the Gauss law constraints remain satisfied in this evolution. 

\end{itemize}

Our simulations are initiated by performing 500 such thermalization
sweeps on a cold configuration. In the physical measurement runs, 
called time trajectories, 
the time evolution according to \eqs\nr{dU_at} and \nr{dE_at} is 
continued for a longer time than in a sweep, typically 
$t^{ }_\rmi{max} = (10 - 100) a$.  
The separate time trajectories, of which there 
are $\sim 10^4_{  } - 10^5_{ }$ 
(cf.\ tables~\ref{table:Gamma_nc2} and \ref{table:Gamma_nc3}), 
are separated by 20 additional sweeps, in order to reduce
autocorrelations. 

%
\subsubsection*{Averaging and error analysis}

The time coordinate $t$ appearing in the measurement of 
\eq\nr{a5Ccl} is really a time difference. 
Given the time-translation invariance of 
our system, this implies that a single trajectory permits for 
many measurements, by considering all time pairings. 
If the $m^\rmi{th}_{ }$ 
trajectory is sampled at times 
$t^{ }_{n} = n a^{ }_t$, 
with $n = 0,...,N-1$, where $(N-1) a^{ }_t = t^{ }_\rmi{max}$ 
sets the maximal time, 
and the average topological charge density is denoted by 
$
 {\bar\chi}^\rmii{$(m)$}_\rmii{$n$} 
 \; \equiv \; 
 \frac{1}{N_s^3}
 \sum_{\vec{x}} 
   \chi^\rmi{(cl)}_{ }(t^{ }_{n},\vec{x})
$, 
then the measurement of \eq\nr{a5Ccl} can be implemented as 
\be
  a^5 C^\rmi{(cl)}_{\S}(t^{ }_j)
  \; = \; 
  \lim_{M\to\infty}
  \frac{N_s^3}{M}
  \,\sum_{m=1}^{M}
  \frac{1}{N-j} 
  \!\sum_{n=0}^{N-j-1}
     \bigl( a^4  {\bar\chi}^\rmii{$(m)$}_\rmii{$n + j$} \bigr) \,   
     \bigl( a^4  {\bar\chi}^\rmii{$(m)$}_\rmii{$n$} \bigr)   
  \;, \la{average}
\ee
where $M$ is the number of trajectories. 
The normalization by $1/(N-j)$ accounts for the fact that 
a single trajectory gives multiple measurements. 

In practice, the number of trajectories is finite, 
and then the limit in \eq\nr{average} cannot be taken. This implies
that the measurement has statistical errors. We estimate them 
with the jackknife method, by dividing the total ensemble of 
trajectories into 
ten separate blocks, and applying the jackknife procedure 
to the blocked measurements.

Given $ a^5 C^\rmi{(cl)}_{\S}(t^{ }_j) $ from \eq\nr{average},  
the Fourier transform from 
\eq\nr{C_w} is taken with Simpson's rule. 
The trapezoidal approximation would yield
\be
  a^4 C^\rmi{(cl)}_{\S}(\omega) 
 \; \simeq \; 
 2 \sum_{j = 0}^{\infty} \frac{a^{ }_t}{a} 
 \cos(\omega t^{ }_j) \, a^5 C^\rmi{(cl)}_{\S}(t^{ }_j) 
 \, 
 \biggl( 1-\frac{\delta^{ }_{j,0}}{2} \biggr)
 \;, \la{a4Ccl_w}
\ee
however given the large amount of data, we have 
taken measurements only at every 5th time step, and therefore
employ a higher-order scheme. Specifically, given 
that the equal-time value $t^{ }_j = 0$ plays an important 
role in the Fourier transform, we have implemented 
a custom routine for the first time step, 
using the knowledge that 
$ C^\rmi{(cl)}_{\S}(t) $ is a symmetric function of $t$. 
The subsequent intervals have been treated with 
quadratic discretization. 

%
\subsubsection*{Finite spatial volume}

\begin{figure}[t]

\hspace*{-0.1cm}
\centerline{%
   \epsfysize=7.5cm\epsfbox{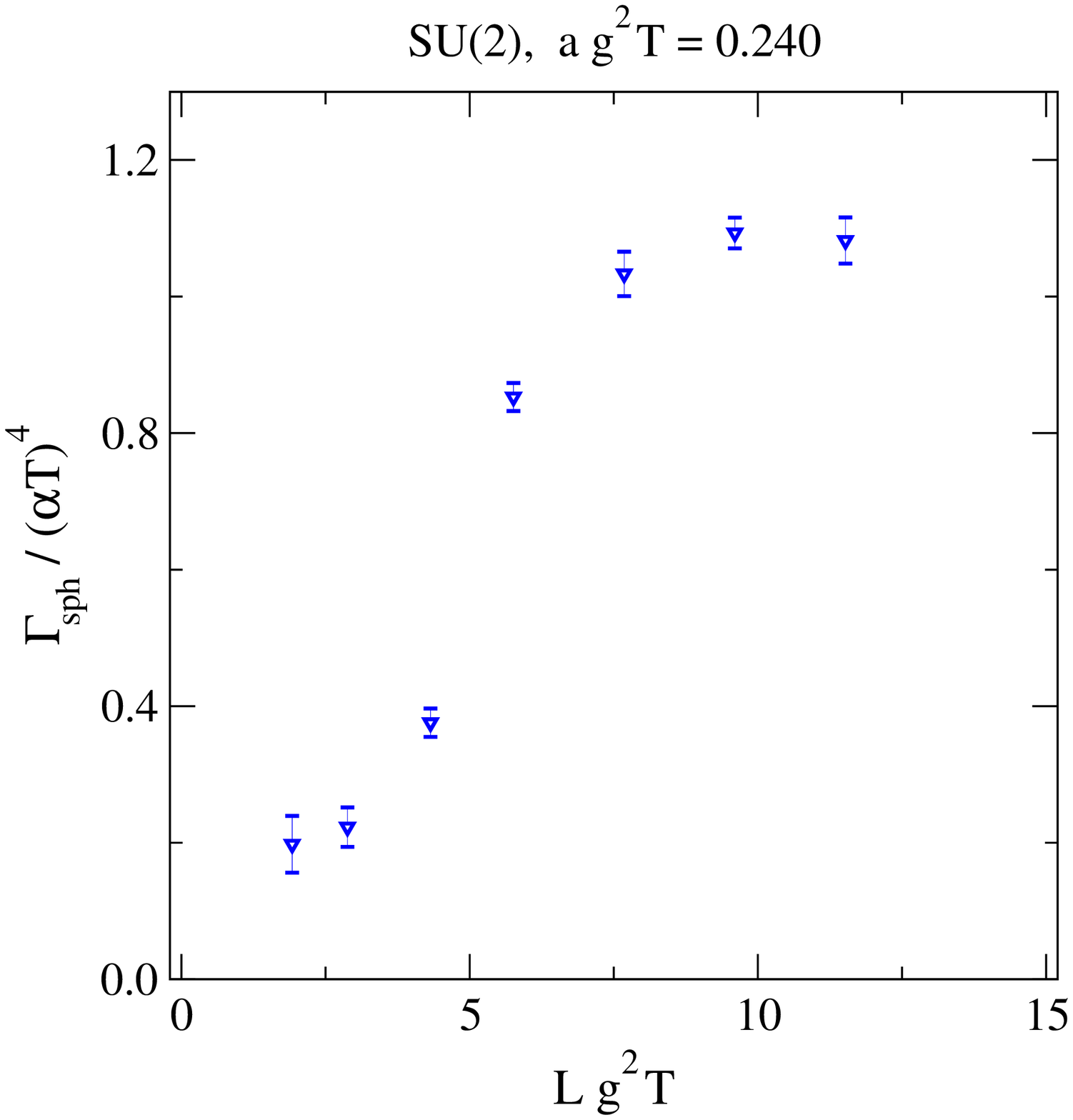}
   ~~~\epsfysize=7.5cm\epsfbox{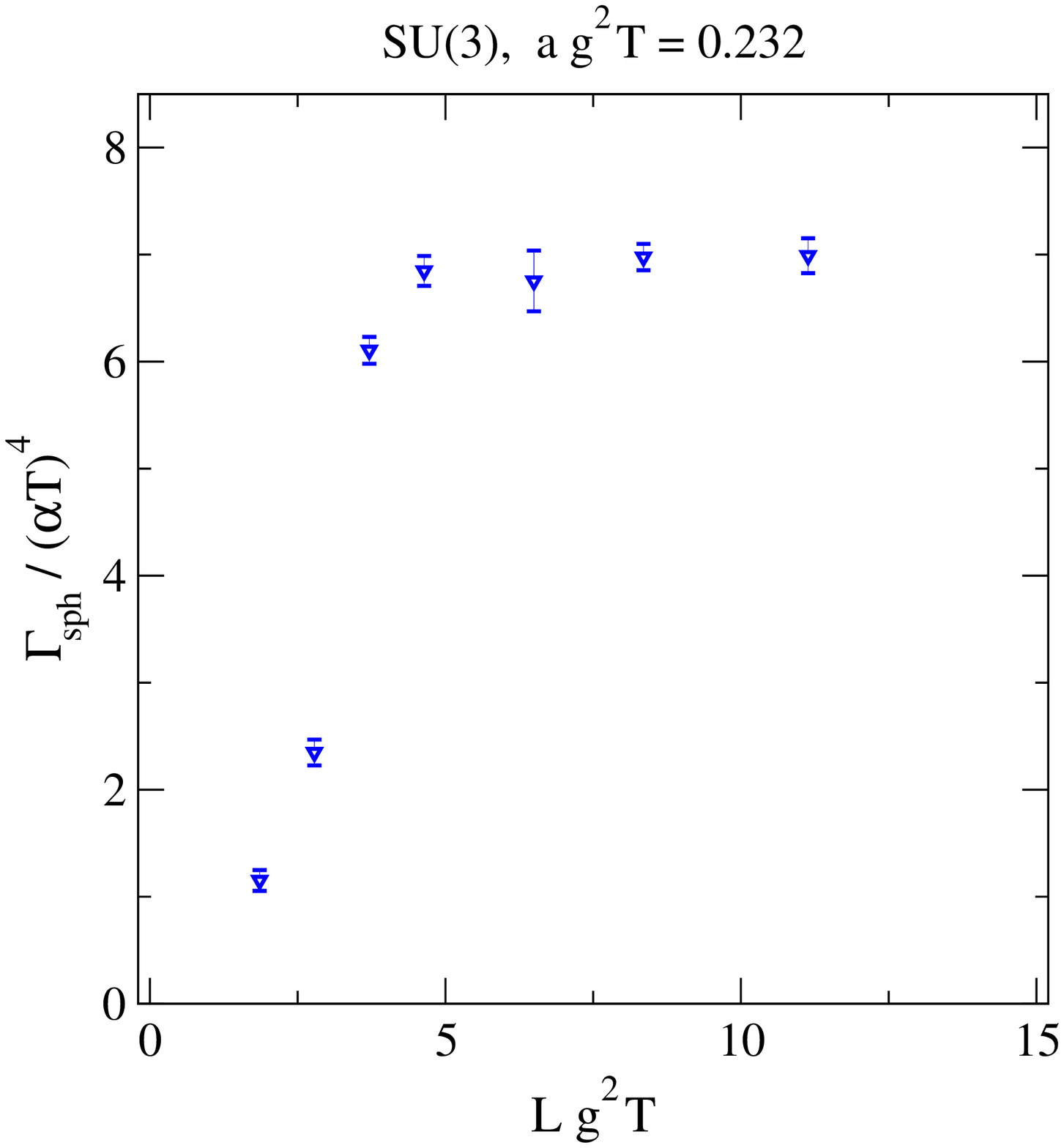}
}

\caption[a]{\small
 Demonstration of the dependence of $\Gamma^{ }_\rmii{\rate}$ 
 from \eq\nr{Gamma_sph} on the spatial extent $L = a N^{ }_s$
 of the box. As expected theoretically~\cite{finiteV}, the 
 volume dependence is exponentially small when $L$ is large
 compared with the colour-magnetic screening scale $\sim 1/(g^2T)$.
 For SU(3), smaller volumes suffice in these units, 
 as thermal glueball masses increase in proportion to $\Nc^{ }$. 
}

\la{fig:finiteV}
\end{figure}

Any simulation takes place in a finite periodic box, 
$V = L^3_{ }$, $L \equiv a N^{ }_s$. Given that non-Abelian
thermal field theory has a mass gap, with the corresponding
confinement scale given by $\sim g^2 T$~\cite{linde}, finite-volume
effects are exponentially small if $L\, g^2 T \gg 1$~\cite{finiteV}. 
However, we need to carry out practical tests to see 
the numerical coefficient on 
the right-hand side of this inequality, so that finite-volume
effects are indeed smaller than statistical errors.  

In \fig\ref{fig:finiteV}, values of $\Gamma^{ }_\rmi{\rate}$
obtained from \eq\nr{Gamma_sph} are shown for a number of box sizes, 
both for $\Nc^{ }=2$ and $\Nc^{ }=3$. For small box sizes, the 
sphaleron rate is reduced (even if with our local definition 
it does not go to zero). As the volume is increased, the rate saturates. 
For SU(2) this happens when $L g^2 T \ge 8.0$, 
for SU(3) when $L g^2 T \ge 5.0$. The box sizes
for our production runs, 
listed in tables~\ref{table:Gamma_nc2} and 
\ref{table:Gamma_nc3}, 
have been chosen so that these inequalities are comfortably satisfied. 

%
\subsubsection*{Finite time extent} 

Another important extent is that of the real-time interval, 
which we denote by $t^{ }_\rmi{max}$. There is less 
theoretical intuition about $t^{ }_\rmi{max}$ than $L$, 
but practical tests are
easier to carry out, as $t^{ }_\rmi{max}$ can be 
varied within a single simulation.  

\begin{figure}[t]

\hspace*{-0.1cm}


\hspace*{-0.1cm}
\centerline{%
   \hspace*{-1mm}\epsfysize=4.2cm\epsfbox{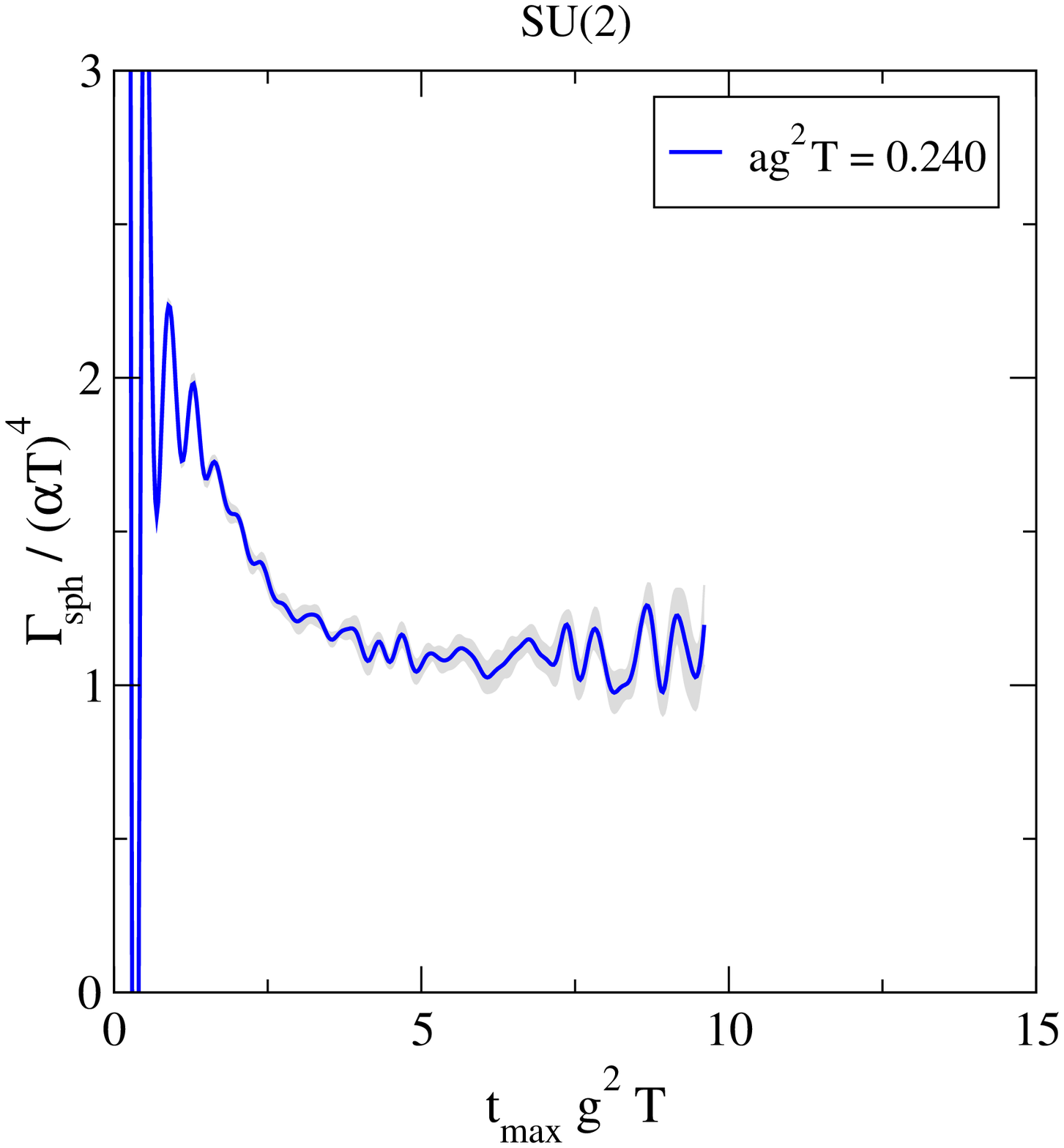}
   \hspace*{-1mm}\epsfysize=4.2cm\epsfbox{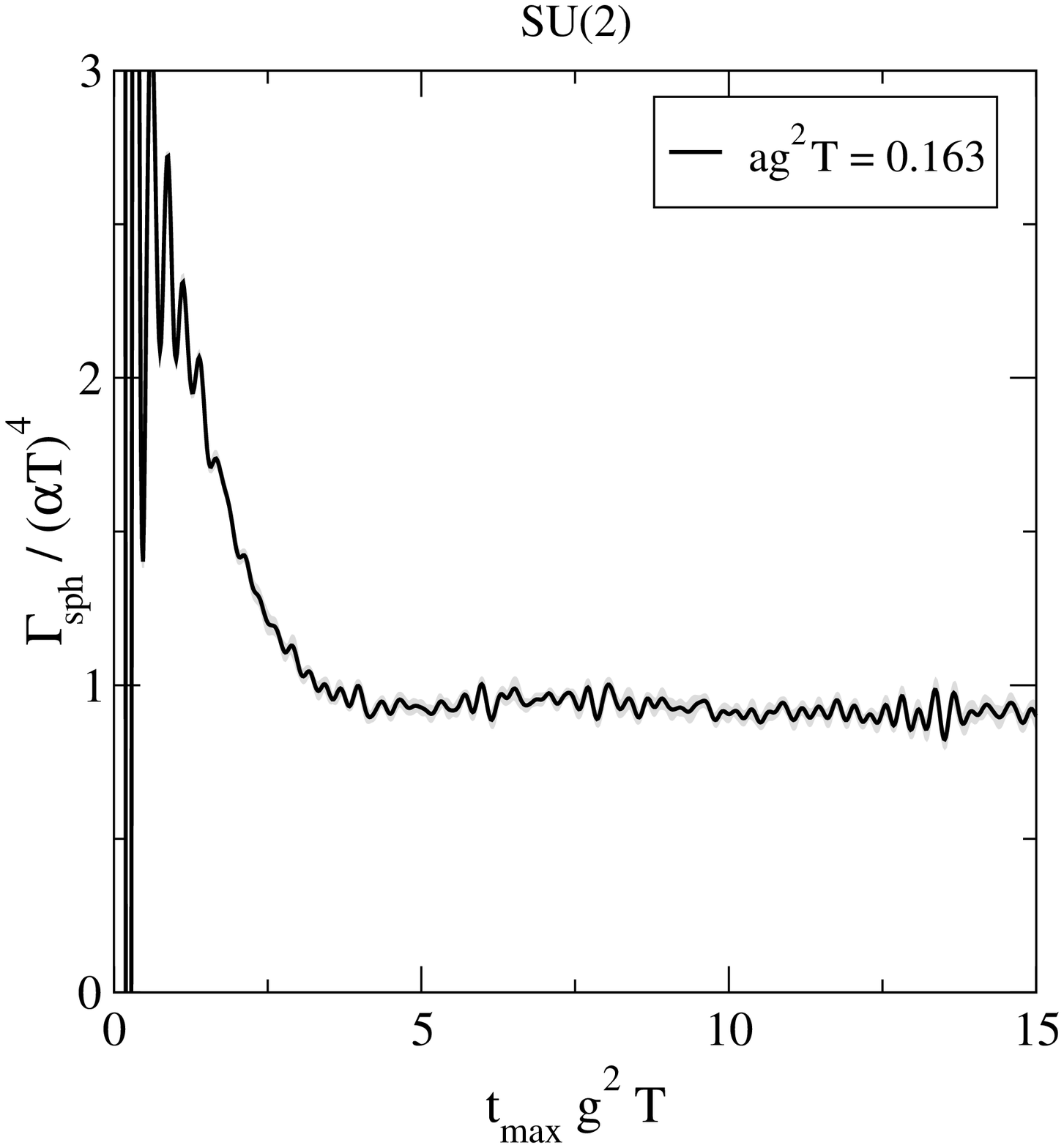}
   \hspace*{-1mm}\epsfysize=4.2cm\epsfbox{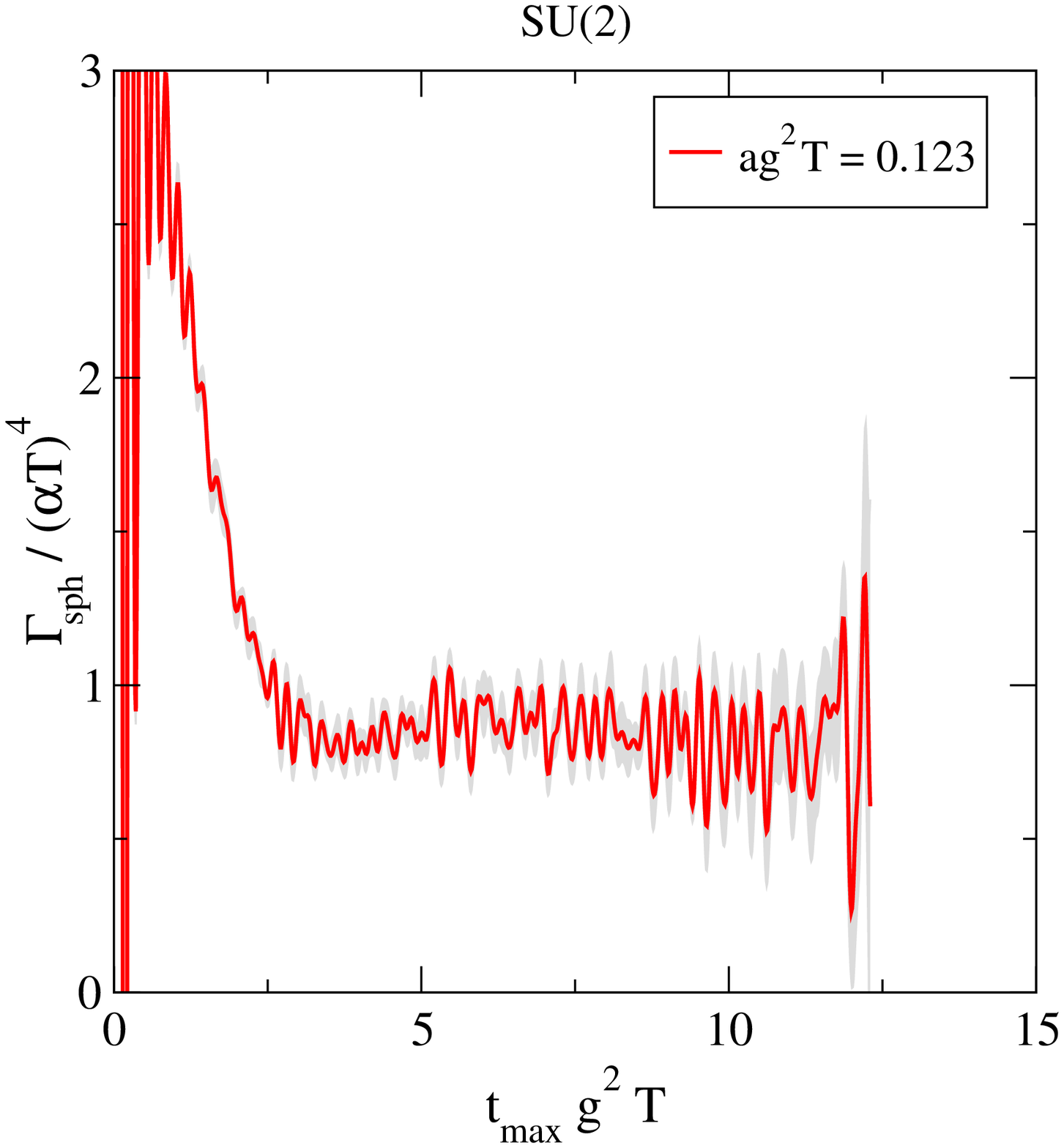}
   \hspace*{-1mm}\epsfysize=4.2cm\epsfbox{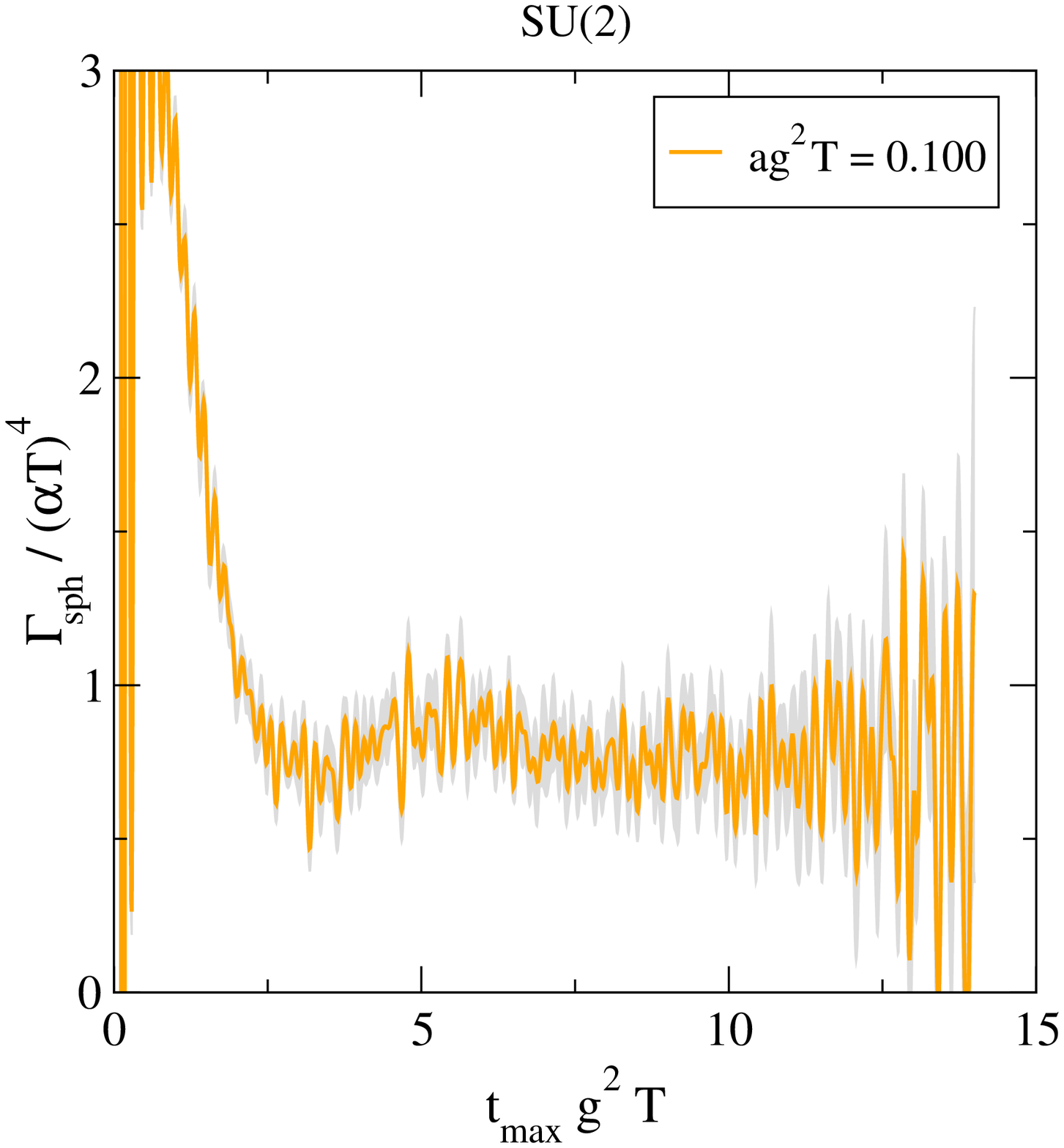}
}

\caption[a]{\small
 Demonstration of the dependence of $\Gamma^{ }_\rmii{\rate}$ 
 from \eq\nr{Gamma_sph} on the time extent $t^{ }_\rmii{max}$. 
 Similarly to the volume dependence
 in \fig\ref{fig:finiteV}, 
 the dependence on $t^{ }_\rmii{max}$ gets stabilized 
 when $t^{ }_\rmii{max}$ is large
 compared with the colour-magnetic scale $\sim 1/(g^2T)$, 
 apart from fast oscillations that average to zero.
}

\la{fig:tmax}
\end{figure}

Examples of sphaleron rates with varying $t^{ }_\rmi{max}$ 
are shown in \fig\ref{fig:tmax}.
Remarkably, we observe a partly similar behaviour as for 
finite-volume effects in \fig\ref{fig:finiteV}, namely a saturation
as a certain value of $t^{ }_\rmi{max}\, g^2 T$ is exceeded, 
which is however now overlayed with rapid oscillations.  

Considering first the saturation point, 
it decreases slightly with decreasing $a g^2 T$. If we associate
the inverse of the saturation point with a frequency scale, we can
then say that there is a frequency scale which increases with 
decreasing $a g^2 T$. Frequency scales could be interpreted as 
threshold energies needed 
to excite a number of glueballs.\footnote{%
  In this language, 
  spectral weight at very small $\omega$ does not 
  correspond to exciting a number of glueballs, 
  but rather to 
  scattering off them. 
 } 
The odd discrete
quantum numbers of $\chi$ suggest that
the glueball ensemble should be odd under the same inversions, 
and that it therefore 
corresponds to a non-perturbative Debye mass~\cite{ay}. 
A constituent Debye mass indeed diverges in classical
lattice gauge theory, cf.\ \eq\nr{mD}. 
A divergence also appears in the height of 
a plateau in
$ C^\rmi{(cl)}_{\S}(\omega) $, 
cf.\ \fig\ref{fig:subtr}.
These observations may provide a rough
explanation for the movement of the saturation point 
as seen in \fig\ref{fig:tmax}.

Turning to the oscillations, we note that they could be interpreted
as a mild ``sign problem'', i.e.\ cancellation between
positive and negative values when integrating over $ C^\rmi{(cl)}_{\S}(t) $
in order to determine 
$
 \lim^{ }_{\,\omega\to 0}  C^\rmi{(cl)}_{\S}(\omega)  
$, 
cf.\ \eq\nr{a4Ccl_w}.
In order to deal with this, 
we have adopted a recipe whereby our measurement of 
$\Gamma^{ }_\rmi{\rate}$ does not originate from a single
fixed $t^{ }_\rmi{max}$, 
but rather measurements in the range 
$t^{ }_\rmi{max}\, g^2 T \in ( \bar\theta \pm \Delta\theta )$.
Within the width $\Delta\theta = 1.0 \; (0.5)$ 
for $\Nc^{ } = 2\; (3)$, 
measurements have been combined as statistical scatter,\footnote{%
 This is supported by the observation that if the noise were 
 uncorrelated and the same number of measurements were carried out
 at each $t^{ }_\rmii{max}$,
 the overall noise magnitude in \fig\ref{fig:tmax} should 
 scale as $\sim\sqrt{t^{ }_\rmii{max}}$. At moderate times,
 a gradual increase can indeed be discerned. 
 } 
whereas the variation $\bar\theta = 5 - 8 \;  (4-6)$ 
for $\Nc^{ } = 2\; (3)$
has been employed for
estimating the systematic uncertainty of this procedure.
The final statistical error is the average of the jackknife errors
from all values of~$\bar\theta$ considered. 
In tables~\ref{table:Gamma_nc2} and \ref{table:Gamma_nc3}, 
the systematic and statistical uncertainties are displayed separately, 
whereas in \fig\ref{fig:comp}, 
they have been combined in quadrature.

Finally, 
we note that a finite value of $t^{ }_\rmi{max}$ 
implies that we have a limited resolution in $\omega$.
Recalling the symmetry of 
$ C^\rmi{(cl)}_{\S}(t) $ under $t\to -t$, 
independent frequencies
should be chosen as multiples of $\pi / t^{ }_\rmi{max}$.

%
\section{Fresh look at the sphaleron rate}
\la{se:cs}

Having established the technical setup, 
we proceed to the results, 
considering first the sphaleron rate, 
\be
 \Gamma^{ }_\rmi{\rate} \; = \; 
 \lim^{ }_{\,\omega\to 0}  C^{ }_{\S}(\omega) 
 \;. \la{Gamma_sph}
\ee
The sphaleron rate 
has previously been measured with Hamiltonian simulations 
in refs.~\cite{mr,mt}, however
with a different definition. While the simulation itself 
proceeded in the same way, for measurements  
the gauge configurations were ``cooled''
towards the solution of classical equations of motion, removing
short-distance fluctuations and arguably turning $\chi$ into
a topological observable.\footnote{%
 Cooling in those studies amounts to the introduction of a fictitious 
 coordinate $\eta$ and the solution of equations of the type
 $\partial^{ }_\eta \Phi = - \delta S^{ }_\rmii{E} / \delta \Phi$, 
 where $S^{ }_\rmii{E}$ denotes 
 the (3-dimensional) Euclidean action and $\Phi$ 
 the ensemble of fields. The initial condition $\Phi(\eta\to 0^+_{ })$
 is taken from the Monte Carlo history,
 whereas the asymptotic value $\Phi(\eta\to \infty)$ satisfies 
 $ \delta S^{ }_\rmii{E} / \delta \Phi = 0$ and is thus
 a solution of equations of motion.   
 }  
Subsequently, $\Gamma^\rmi{cooled}_\rmi{\rate}$
was extracted as a diffusion coefficient related to the movement of
the cooled Chern-Simons number. 
We note that even though physically intuitive,
such a definition does not lend itself to an operator definition that
emerges naturally from quantum field theory; a cooled spectral
function is not the same object that affects, e.g.,
the motion of an axion field coupled to a local pseudoscalar operator.
In other words, 
cooling changes the {\em expectation value} of  
$\Gamma^{ }_\rmi{\rate}$, not only its variance.

\begin{table}[t]

\hspace*{-0.1cm}
\begin{minipage}[c]{15.2cm}
\small{
\begin{center}
\vspace*{-4mm}
\begin{tabular}{cccccccc} 
 \multicolumn{8}{c}{$\Nc^{ } = 2$}
 \\[2mm]
 \hline 
 \\[-4mm]
 $ \beta^{ }_\rmii{L}$ & 
 $ \!\!\! a^{ }_\rmi{bare}\, g^2 T \!\!\! $ &
 $ N^{ }_s $ &
 $ L\, g^2 T $ &
 $ t^{ }_\rmi{max}\, g^2 T $ &
   \# traj.\ &
 $ \Gamma_\rmi{\rate}^\rmi{ } / (\alpha T)^4 $ &  
 $ \!\!\!\Gamma_\rmi{\rate\!\mbox{\cite{mr}}}^\rmi{cooled}  / (\alpha T)^4 $
 \\[2mm]
 \hline
 \\[-4mm] 
 4.63... & 0.863 & 10 & 8.63 & $\le 9.49$ & 50000 
 & 0.885(0.010)$^{ }_\rmii{syst}$(0.008)$^{ }_\rmii{stat}$ & 3.82(0.14)
  \\
 6.60... & 0.606 & 16 & 9.70 & $\le 9.09$ & 100000 
 & 1.061(0.008)$^{ }_\rmii{syst}$(0.011)$^{ }_\rmii{stat}$ & 2.69(0.08)
  \\
 8.64... & 0.463 & 20 & 9.26 & $\le 9.26$ & 57205 
 & 1.129(0.010)$^{ }_\rmii{syst}$(0.013)$^{ }_\rmii{stat}$ & 2.28(0.04)
  \\
 12.7... & 0.314 & 32 & 10.0 & $\le 12.6$ & 50000 
 & 1.128(0.025)$^{ }_\rmii{syst}$(0.011)$^{ }_\rmii{stat}$ & 1.64(0.04)
  \\
 13.5... & 0.295 & --- & --- & --- & --- 
 & --- & 1.53(0.04)
  \\
 16.6... & 0.240 & 40 & 9.60 & $\le 9.60$ & 46580 
 & 1.093(0.008)$^{ }_\rmii{syst}$(0.021)$^{ }_\rmii{stat}$ & 1.31(0.05)
  \\
 24.5... & 0.163 & 60 & 9.78 & $\le 26.1$ & 78470 
 & 0.944(0.012)$^{ }_\rmii{syst}$(0.004)$^{ }_\rmii{stat}$ & 0.95(0.03)
  \\
 32.5... & 0.123 & 80 & 9.84 & $\le 12.3$ & 59652 
 & 0.870(0.023)$^{ }_\rmii{syst}$(0.012)$^{ }_\rmii{stat}$ & 0.68(0.03)
  \\
 40.0... & 0.100 & 96 & 9.60 & $\le 14.0$ & 41132 
 & 0.816(0.059)$^{ }_\rmii{syst}$(0.010)$^{ }_\rmii{stat}$ &   --- 
  \\
 53.3... & 0.075 & 128 & 9.60 & $\le 12.0$ & 33373 
 & 0.742(0.112)$^{ }_\rmii{syst}$(0.015)$^{ }_\rmii{stat}$ &   --- 
  \\
 \hline 
\end{tabular} 
\end{center}
}
\end{minipage}

\vspace*{3mm}

\caption[a]{\small
     Our results for the sphaleron rate for $\Nc^{ } = 2$,
     compared with
     ``cooled'' rates
     from table~2 of ref.~\cite{mr},
     however we have undone the improvement 
     employed in the latter study,
     by making use of \eq\nr{impr}. 
     The values
     of $ a^{ }_\rmii{bare}\, g^2 T $ are treated as 
     ``exact'', whereas
     digits have been truncated from 
     the values of $\beta^{ }_\rmii{L}$ shown (cf.\ \eq\nr{betaL}).
     These results are illustrated in \fig\ref{fig:comp}.
}

\la{table:Gamma_nc2}
\end{table}

\begin{table}[t]

\hspace*{-0.1cm}
\begin{minipage}[c]{15.2cm}
\small{
\begin{center}
\vspace*{-4mm}
\begin{tabular}{cccccccc} 
 \multicolumn{8}{c}{$\Nc^{ } = 3$}
 \\[2mm]
 \hline 
 \\[-4mm]
 $ \beta^{ }_\rmii{L}$ & 
 $ \!\!\! a^{ }_\rmi{bare}\, g^2 T \!\!\! $ &
 $ N^{ }_s $ & 
 $ L\, g^2 T $ &
 $ t^{ }_\rmi{max}\, g^2 T $ &
   \# traj.\ &
 $ \Gamma_\rmi{\rate}^\rmi{ } /(\alpha T)^4 $ & 
 $ \!\!\!\Gamma_\rmi{\rate\!\mbox{\cite{mt}}}^\rmi{cooled} /(\alpha T)^4 $
 \\[2mm]
 \hline
 \\[-4mm] 
 9.35... & 0.642 & 20 & 12.8 & $\le 10.0$ & 78775  
 & 5.054(0.037)$^{ }_\rmii{syst}$(0.024)$^{ }_\rmii{stat}$ & 41.6(0.7) 
  \\
 10.8... & 0.553 & 20 & 11.1 & $\le 10.1$ & 41738  
 & 5.575(0.038)$^{ }_\rmii{syst}$(0.040)$^{ }_\rmii{stat}$ & 29.6(0.4) 
  \\
 12.3... & 0.488 & 20 & 9.76 & $\le 10.2$ & 32207 
 & 6.101(0.061)$^{ }_\rmii{syst}$(0.055)$^{ }_\rmii{stat}$ & 24.1(0.5) 
  \\
 13.8... & 0.433 & 20 & 8.66 & $\le 10.0$ & 59666 
 & 6.434(0.011)$^{ }_\rmii{syst}$(0.045)$^{ }_\rmii{stat}$ & 20.3(0.5)
  \\
 16.8... & 0.356 & 20 & 7.12 & $\le 10.0$ & 54168  
 & 6.764(0.022)$^{ }_\rmii{syst}$(0.025)$^{ }_\rmii{stat}$ & 16.5(0.3)
  \\
 19.8... & 0.302 & 20 & 6.04 & $\le 10.3$ & 58157 
 & 6.917(0.037)$^{ }_\rmii{syst}$(0.058)$^{ }_\rmii{stat}$ & 13.6(0.3)
  \\
 25.8... & 0.232 & 36 & 8.35 & $\le 8.12$ & 48400 
 & 6.977(0.084)$^{ }_\rmii{syst}$(0.090)$^{ }_\rmii{stat}$ & 11.4(0.2)
  \\
 31.9... & 0.188 & 40 & 7.52 & $\le 8.46$ & 63657 
 & 6.641(0.026)$^{ }_\rmii{syst}$(0.069)$^{ }_\rmii{stat}$ & 9.9(0.3) 
  \\
 48.0... & 0.125 & 60 & 7.50 & $\le 8.75$ & 55072 
 & 5.780(0.158)$^{ }_\rmii{syst}$(0.086)$^{ }_\rmii{stat}$ & --- 
  \\
 60.0... & 0.100 & 80 & 8.00 & $\le 8.00$ & 48968 
 & 5.300(0.227)$^{ }_\rmii{syst}$(0.086)$^{ }_\rmii{stat}$ & --- 
  \\
 80.0... & 0.075 & 96 & 7.20 & $\le 8.25$ & 35815 
 & 4.903(0.393)$^{ }_\rmii{syst}$(0.130)$^{ }_\rmii{stat}$ & --- 
  \\
 \hline 
\end{tabular} 
\end{center}
}
\end{minipage}

\vspace*{3mm}

\caption[a]{\small
     Like table~\ref{table:Gamma_nc2} but for $\Nc^{ } = 3$.
     In this case the cooled rates originate 
     from table~3 of ref.~\cite{mt},
     after undoing the improvement 
     employed in that study.
}

\la{table:Gamma_nc3}
\end{table}

Another feature in refs.~\cite{mr,mt} is 
$\rmO(a)$ improvement of the parameters used. 
The idea is that the coefficient
in front of the Hamiltonian in \eq\nr{Z_cl} is replaced through~\cite{mt} 
\be
 \frac{1}{a g^2 T}  
 \; \longrightarrow \; 
 \frac{1}{a^{ }_\rmi{bare}\, g^2 T} \; \equiv \; 
 \frac{1}{a^{ }_\rmi{impr}\, g^2 T} 
 + \Nc \, \biggl(
             0.12084 - \frac{1}{6 \Nc^2} 
          \biggr) 
 \;. \la{impr}
\ee
The lattice spacing is conventionally reparametrized as 
\be
 \beta^{ }_{\rmii{L}} \; \equiv \;
 \frac{2\Nc^{ }}{ a^{ }_\rmi{bare}\, g^2 T }
 \;. 
 \la{betaL} 
\ee
For us improvement is unnecessary and complicates 
the discussion, whence we re-tabulate
the results from refs.~\cite{mr,mt} in 
the last columns of tables~\ref{table:Gamma_nc2} 
and~\ref{table:Gamma_nc3}
without it.
In our notation, 
\be
 ag^2 T \;\equiv\; a^{ }_\rmi{bare}\, g^2 T
 \;.
\ee

\begin{figure}[t]

\hspace*{-0.1cm}
\centerline{%
   \epsfysize=7.5cm\epsfbox{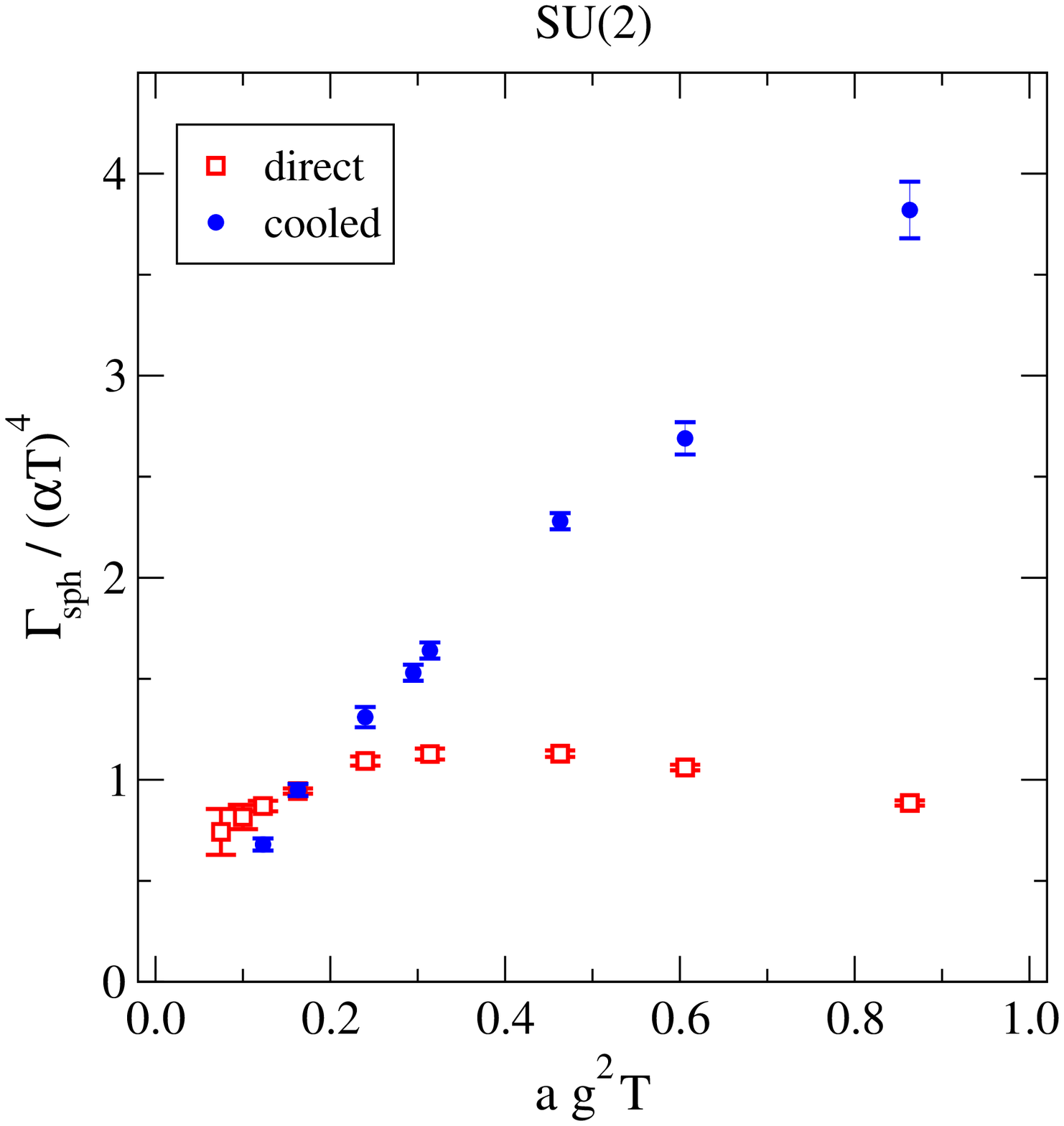}
   ~~~\epsfysize=7.5cm\epsfbox{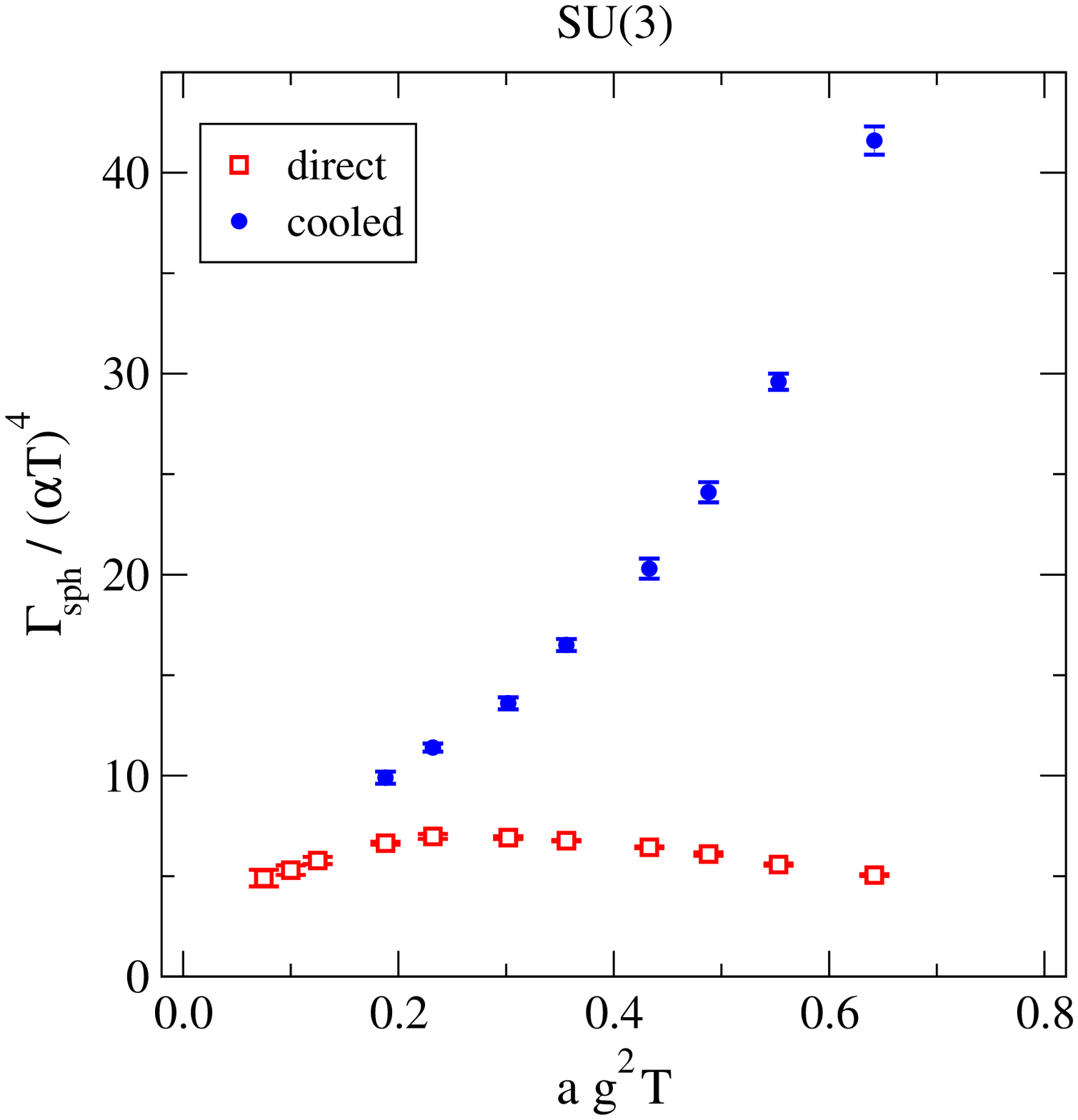}
}

\caption[a]{\small
 Comparison of our $\Gamma^{ }_\rmii{\rate}$ 
 from \eq\nr{Gamma_sph} (``direct'') against literature results
 employing a different definition (``cooled'')~\cite{mr,mt}.
 The data are as given 
 in tables~\ref{table:Gamma_nc2} and~\ref{table:Gamma_nc3},
 with errors combined in quadrature. 
 The results do not agree quantitatively, but both approach 
 zero in  the scaling regime $ag^2 T \ll 0.2$, 
 with an approximately linear slope,
 modified by a complicated logarithmic dependence. 
}

\la{fig:comp}
\end{figure}

Our results for $\Gamma^{ }_\rmi{\rate}$ are listed
in tables~\ref{table:Gamma_nc2} and~\ref{table:Gamma_nc3}, and 
shown in \fig\ref{fig:comp}. The pattern observed is simple
to state: at coarse lattices, $ag^2 T \gg 0.2$, our results
are clearly below the cooled ones. 
In contrast, as we go to 
$ag^2T \ll 0.2$, two things happen simultaneously: 

\begin{itemize}

\item[(i)]
Our results for $\Gamma^{ }_\rmi{\rate}$ are of the same order
of magnitude as 
$\Gamma^\rmi{cooled}_\rmi{\rate}$. 
The reason should be that in this regime a scale hierarchy
sets in, 
and the universal IR dynamics responsible for the sphaleron rate
takes over from non-universal UV fluctuations. 
The results are not exactly the same, however. 
This need not be
surprising, as the IR and UV fluctuations do not 
decouple from each other in a non-renormalizable theory, 
and as it is difficult to specify which precise UV 
fluctuations have been removed by the cooling. 

\item[(ii)]
The sphaleron rate $\Gamma^{ }_\rmi{\rate}$ displays
approximately linear scaling with $ag^2T$, as 
expected from \eq\nr{ansatz_intercept}. 
This is a key feature of the universal IR dynamics. 
It has, however, been modified by a complicated
(unknown but likely 
non-polynomial) logarithmic dependence, originating from the 
non-decoupling of the IR and UV fluctuations. 

\end{itemize}

We refer to the regime $ag^2 T \ll 0.2 $,
in which these observations apply, as the {\em scaling regime}. 
It is only 
in the scaling regime that classical lattice gauge theory 
can be employed for extracting physical information, 
and therefore we focus on it in the following. 

%
\section{Estimate of the shape of the spectral function}
\la{se:shape} 

We finally turn to the $\omega$-dependence
of $C^\rmi{(cl)}_{\S}(\omega)$. 
In \fig\ref{fig:bare}, results 
normalized according to \eq\nr{normalization} are plotted 
at a few small lattice spacings, where we are in the scaling regime. 
The results are compared with a leading-order perturbative 
evaluation, detailed in appendix~A. For a better resolution, 
these results
are replotted in \fig\ref{fig:subtr}, after the subtraction of 
the perturbative part and a conversion of the frequency scale
into units of $g^2 T$. 

\begin{figure}[t]

\hspace*{-0.1cm}
\centerline{%
   \epsfysize=7.5cm\epsfbox{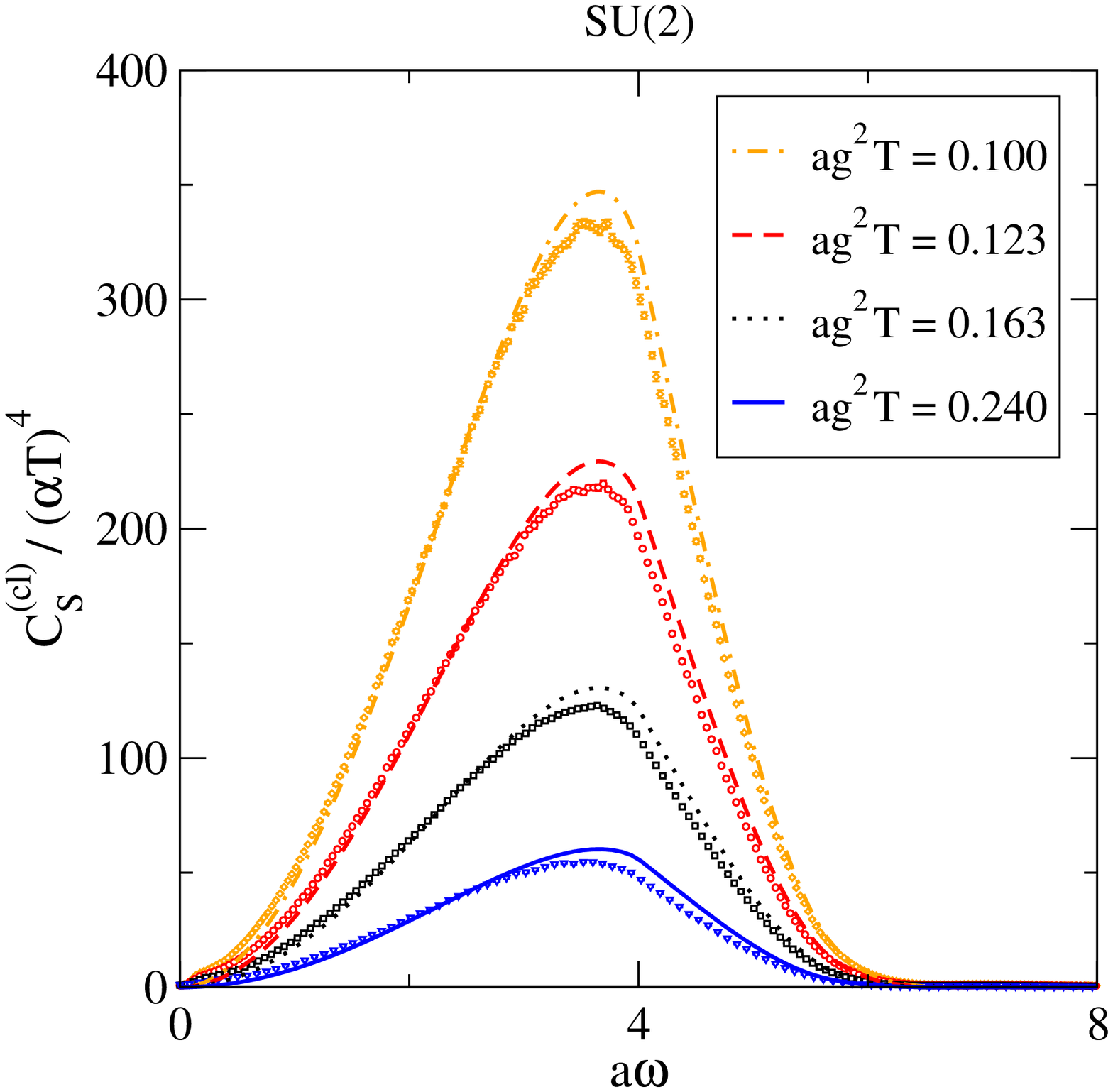}
   ~~~\epsfysize=7.5cm\epsfbox{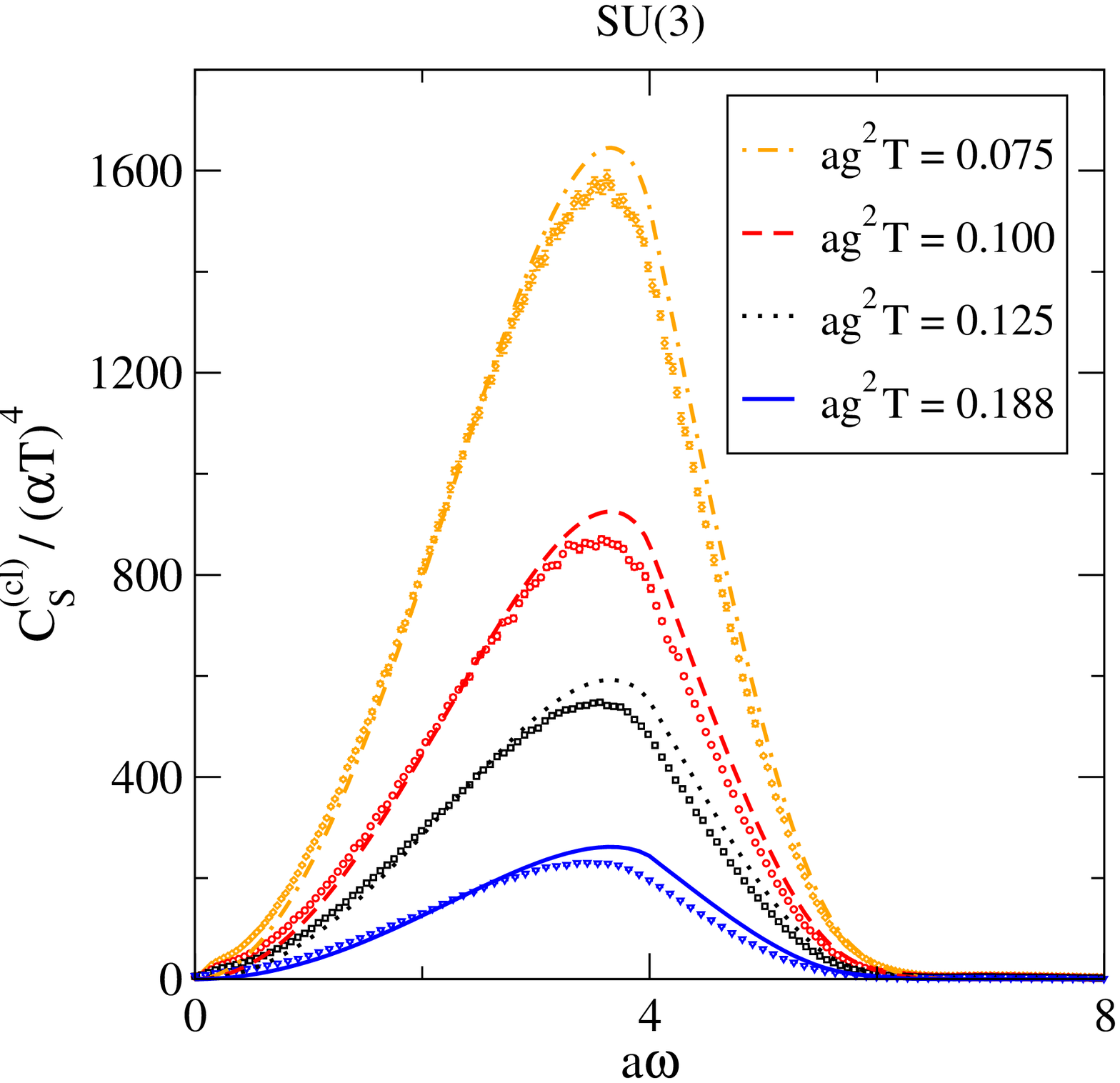}
}

\caption[a]{\small
 Lattice data (open symbols) for 
 $
  C^\rmi{(cl)}_{\S}(\omega) / (\alpha T)^4
 $, 
 with values of $a g^2 T$ chosen from the scaling regime
 (cf.\ \fig\ref{fig:comp}), 
 compared with the perturbative contribution
 from appendix~A (lines). 
 The latter is a fairly good approximation in the UV domain 
 $\omega \sim 1/a$, with the small discrepancy conceivably
 due to next-to-leading order corrections.
 Our physics conclusions come instead from the IR  
 domain $a\omega = a g^2 T \times \omega / (g^2 T) \lsim 1.3$, 
 cf.\ \fig\ref{fig:subtr}, so we need to zoom very close
 to the origin to extract them.
}

\la{fig:bare}
\end{figure}

\begin{figure}[t]

\hspace*{-0.1cm}
\centerline{%
   \epsfysize=7.5cm\epsfbox{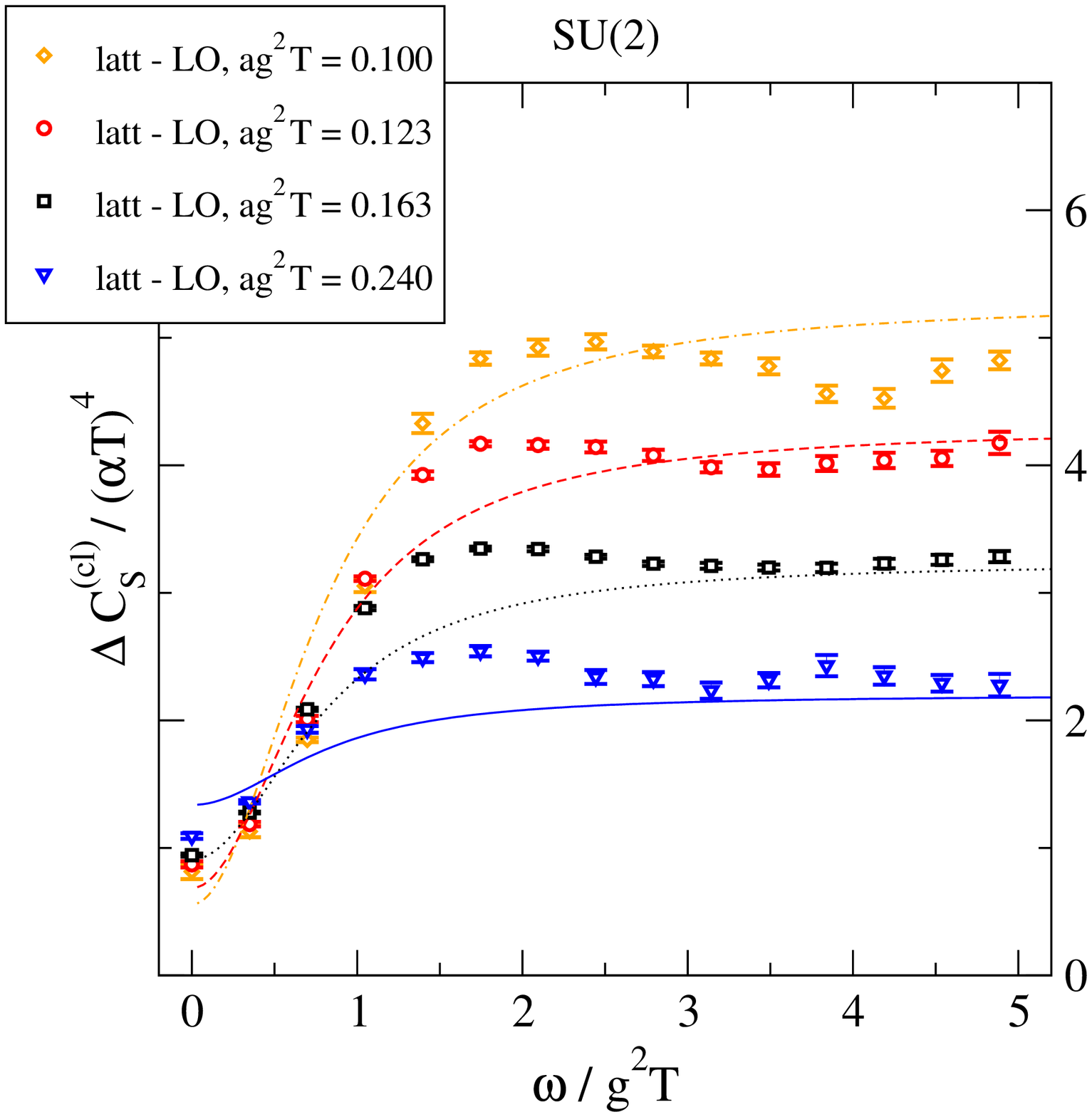}
   ~~~\epsfysize=7.5cm\epsfbox{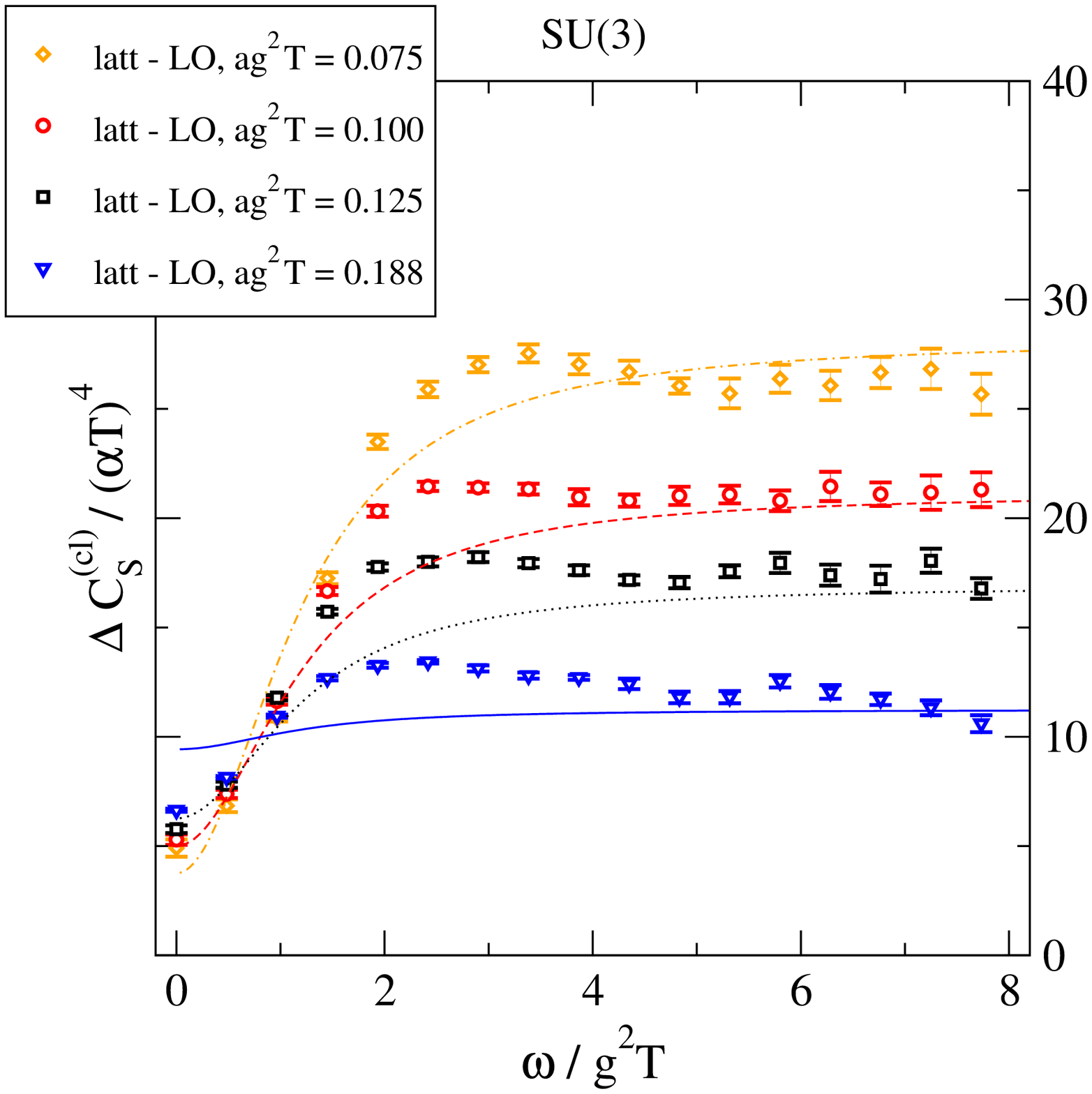}
}

\caption[a]{\small
 Lattice data for 
 $
  \Delta C^\rmi{(cl)}_{\S}(\omega) / (\alpha T)^4
 $
 as a function of $\omega / (g^2 T)$, 
 after the subtraction of the 
 perturbative contribution from appendix~A, 
 computed up to leading order (``LO'').
 The lines indicate a qualitative
 representation of the subtracted data, 
 according to \eqs\nr{model} and \nr{fit_nc3}.
}

\la{fig:subtr}
\end{figure}

The results in \fig\ref{fig:subtr} represent
an IR contribution, whose shape we would like to understand.  
Generalizing on ref.~\cite{warm}, the IR part 
may be parametrized as 
\ba
 \Delta \rho(\omega) \big|^{ }_\rmi{cont} 
  & \equiv & 
 \rho(\omega) \big|^{ }_\rmi{cont}
 \; - \; 
 \rho^{ }_\rmiii{UV}(\omega) \big|^{ }_\rmi{cont} 
 \\[2mm]
 & \stackrel{ |\omega| \,\ll\, g T }{\simeq} & 
 \omega \Upsilon^{ }_{\rmiii{IR}}
 \mathcal{F}
 \biggl( \frac{\omega^2}{\omega_\rmiii{IR}^2},
         \frac{\omega^2}{\omega_\rmiii{M}^2},... \biggr)
 \;, \\[2mm]
 \Upsilon^{ }_\rmiii{IR} \big|^{ }_\rmi{cont} 
 & \equiv & \kappa\, \alpha^5 \Nc^3 (\Nc^2 - 1 ) T^3
 \;, \\[2mm]
 \lim_{\omega\to 0}\mathcal{F}
 & \equiv & 1
 \;. \la{cont_params} \hspace*{6mm}
\ea
Here 
$ 
 \rho^{ }_\rmiii{UV}(\omega)
$
denotes a perturbative contribution, 
applicable to large frequencies 
$
 |\omega| \gsim gT
$.
The scaling factor $\Nc^2 - 1$ in $\Upsilon^{ }_{\rmiii{IR}}$ conforms with 
ref.~\cite{mt}, and is reproduced by a perturbative 
computation, cf.\ \eq\nr{Ccl1}, where it appears as 
the dimension of the adjoint representation,~$\dA^{ }$.
According to \eq\nr{relation2}, the corresponding
statistical correlator reads 
\be
 \Delta C^{ }_{\S}(\omega) \bigg|^{ }_\rmi{cont}
 \;\stackrel{ |\omega| \,\ll\, g T }{\simeq}\;
 \underbrace{ 2 T \Upsilon^{ }_{\rmiii{IR}} }_{ \Gamma^{ }_\rmi{\rate} } 
 \mathcal{F}
 \biggl( \frac{\omega^2}{\omega_\rmiii{IR}^2},
         \frac{\omega^2}{\omega_\rmiii{M}^2},... \biggr)
 \;. \la{cont_ansatz}
\ee

The parameters $\omega^{ }_i$ in \eq\nr{cont_ansatz}
represent 
frequency scales, the most IR of which is of order 
\be
 \omega^{ }_\rmiii{IR} \big|^{ }_\rmi{cont} 
 \simeq c^{ }_\rmiii{IR} \,
 \alpha^2 \Nc^2 T 
 \;, \la{Delta_1}
\ee
for a suitable coefficient $ c^{ }_\rmiii{IR} $.
But there are also other frequency scales, for instance the ``magnetic''
or Linde scale~\cite{linde}, 
required for exciting non-perturbative glueball-like states, 
\be
 \omega^{ }_\rmiii{M} \big|^{ }_\rmi{cont} 
 \simeq c^{ }_\rmiii{M}\,
 \alpha \Nc^{ } T
 \;, \la{Delta_2}
\ee
where $ c^{ }_\rmiii{M} $ is another coefficient of order unity. 

Now, we cannot directly convert \eq\nr{cont_ansatz} 
to lattice units, as one combination
$\sim\alpha^2 \Nc^2 T$, 
closely related to $\omega^{ }_\rmiii{IR}$, will be modified 
on the lattice into $\sim \alpha^2 \Nc^2 T^2 a$.
It is understood, however, 
that modulo an unknown logarithmic dependence, 
this frequency scale is inversely proportional to 
a Debye mass scale $\mD^2$~\cite{clgt4}, where 
\be
 m^2_\rmii{D,cont}
 \stackrel{\rmiii{\Nf=0}}{ =} \frac{\Nc^{ }\, g^2 T^2 }{3} 
 \;, \quad
 m^2_\rmii{D,latt} 
 = 2 \Nc^{ }\, g^2_{ } T \times \frac{\Sigma^{ }_{v^1}}{4\pi a}
 \;. \la{mD}
\ee
The coefficient here has the numerical value
$ 
 \Sigma^{ }_{v^1} \approx 2.1498783949
$, 
where we adopted the more general
notation of ref.~\cite{Mlattice}.
Therefore, continuum estimates can be converted
into lattice estimates, by rescaling them with appropriate powers
of the Debye mass squared~\cite{clgt4}. 

Concretely, multiplying the intercept $\Gamma^{ }_\rmi{\rate}$ by 
$
 m^2_\rmii{D,cont} / 
 m^2_\rmii{D,latt} 
$, 
we expect the $\omega\to 0$ limit to take the form 
\be
 \lim_{\omega\to 0} 
 \frac{ \Delta C^\rmi{(cl)}_{\S}(\omega) }{  (\alpha T)^4 }
 \bigg|^{ }_\rmi{latt}
 \;\simeq\; 
 \frac{ m^2_\rmii{D,cont} }{ m^2_\rmii{D,latt} }
 \biggl\{ 
 \lim_{\omega\to 0} 
 \frac{ \Delta C^\rmi{ }_{\S}(\omega) }{  (\alpha T)^4 }
 \bigg|^{ }_\rmi{cont}
 \biggr\} 
 \;\simeq\; 
 \frac{\kappa\, \Nc^3 (\Nc^2 - 1) a g^2 T}{3 \Sigma^{ }_{v^1}}
 \;. \la{ansatz_intercept}
\ee
Scaling similarly the width from \eq\nr{Delta_1}, we find
\be
 \omega^{ }_\rmiii{IR} \big|^{ }_\rmi{latt}
 \; \simeq \;
 \frac{ m^2_\rmii{D,cont} }{ m^2_\rmii{D,latt} }
 \, \omega^{ }_\rmiii{IR} \big|^{ }_\rmi{latt}
 \; \simeq  \; 
 \frac{c^{ }_\rmiii{IR}\, a (g^2 T \Nc^{ })^2}{24 \pi \Sigma^{ }_{v^1}} 
 \;. \la{ansatz_Delta}
\ee
In contrast, the magnetic scale $\omega^{ }_\rmiii{M}$ from \eq\nr{Delta_2} 
should remain intact on the lattice. 

Taking a look at \fig\ref{fig:subtr}, we observe that the intercept
at $\omega\to 0 $ scales towards zero as the lattice spacing is 
decreased, in accordance with \eq\nr{ansatz_intercept}. But the 
curves cross each other as $\omega$ is increased, 
and the plateau obtained at $\omega\gsim g^2 T$ scales in the
opposite direction, as $\sim 1 / (ag^2 T)$. This suggests that 
\be
 \mathcal{F} 
 \;\; \stackrel{\omega\ll 
                \omega^{ }_\rmiii{IR},\,\omega^{ }_\rmiii{M}
                }{\equiv}
 \;\; 1
 \;, \quad 
 \mathcal{F} 
 \;\; \stackrel{\omega\gg
                \omega^{ }_\rmiii{IR},\,\omega^{ }_\rmiii{M}
               }{\propto}
 \;\; \frac{\omega^{2}_\rmiii{M}}{\omega^{2}_\rmiii{IR}}
 \;. \la{observation}
\ee
A simple ansatz satisfying \eq\nr{observation} reads
\be
 \mathcal{F} \; \simeq  \;
   {\frac{1 + \frac{\omega^2}{\omega_\rmiii{IR}^2} }
         {1 + \frac{\omega^2}{\omega_\rmiii{M}^2}  }
   }
 \;, \la{F_ansatz_cont}
\ee 
which gives a 3-parameter model 
($\kappa, c^{ }_\rmiii{IR}, c^{ }_\rmiii{M}$)
with which to compare the lattice data, {\it viz.} 
\be
 \frac{\Delta C^\rmi{(cl)}_{\S}(\omega) }{  (\alpha T)^4 }
 \bigg|^{ }_\rmi{latt}
 \;\simeq\; 
 \frac{\kappa\, \Nc^3 (\Nc^2 - 1) a g^2 T}{3 \Sigma^{ }_{v^1}}
 \;
     {\frac{1 + \bigl(\frac{\omega}{g^2T}\bigr)^2
                 \Bigl(\frac{24\pi \Sigma_{v^1}}
                  {c^{ }_\rmiii{IR} a g^2_{ }T \Nc^2} \Bigr)^2 
            }
            {1 + \bigl(\frac{\omega}{g^2T}\bigr)^2
                  \Bigl(\frac{4\pi}
                  {c^{ }_\rmiii{M} \Nc^{ }} \Bigr)^2
            }} 
 \;. \la{model}
\ee

As demonstrated by 
the lines in \fig\ref{fig:subtr},
the data can indeed be 
represented with \eq\nr{model}
on a qualitative level. 
The representation is not perfect,  
with perhaps the main 
deficiency that the threshold between the dip at 
small $\omega$ and the plateau at large $\omega$
is somewhat too shallow.  
Overlooking this feature, 
the parameters can be chosen as\hspace*{0.2mm}\footnote{%
  Technically, the parameters have been found 
  with a $\chi^2$ minimization of SU(2) data, 
  weighting however the important intercepts at $\omega\to 0$
  with an additional factor $\sim 10^3_{ }$,
  but we refrain from citing $\chi^2_{ }$ for such a  
  fit, as the representation is a phenomenological one, 
  rather than based on an {\em a priori} justified ansatz. 
  The SU(3) data was not used
  for tuning the coefficients,
  but it is nevertheless represented reasonably well by them. 
 }
\ba
 && 
 \kappa \,\simeq\, 1.5  \;, \quad
 c^{ }_\rmiii{IR} \,\simeq\, 106 \;, \quad
 c^{ }_\rmiii{M} \,\simeq\, 5.1 
 \;. \la{fit_nc3}
\ea
On the qualitative level, the same parameters 
$\kappa$, $c^{ }_\rmiii{IR}$, and $c^{ }_\rmiii{M}$ 
can be used in the continuum
estimates of \eqs\nr{cont_params}--\nr{Delta_2}, 
cf.\ \eq\nr{final_cont}.

\vspace*{3mm}

We conclude this section by recalling that 
an alternative view on the 
conversion from lattice to continuum units is that, 
following refs.~\cite{clgt4,mt}, 
we can equate the two Debye masses in \eq\nr{mD}, yielding
\be
 \frac{g^2}{4\pi}
 \; \simeq \; \frac{ag^2 T}{6 \Sigma^{ }_{v^1}}
 \;. \la{match}
\ee
This implies that a study with $a g^2 T \simeq 0.20$
can ``simulate'' a world with $\alpha \simeq 0.015$, 
as was the case during the reheating period
as studied in ref.~\cite{warm}. The corresponding IR part of 
the spectral shape could be extracted from \fig\ref{fig:subtr}, 
directly in physical units.

%
\section{Conclusions and outlook}
\la{se:concl}

Several frequency scales 
characterize the dynamics of thermalized 
non-Abelian fields (cf., e.g.,\ ref.~\cite{ay_t}). 
As reviewed at the end of \se\ref{se:formulation}, 
the most IR among them, playing an essential role for 
sphaleron dynamics, is of order 
$\omega^{ }_\rmiii{IR}
\sim (\alpha \Nc^{ }T )^3 / m_\rmiii{E}^2 \sim \alpha^2 \Nc^2 T / \pi$, 
where $m_\rmiii{E}^{2} \sim g^2 \Nc^{ } T^2$ 
is the electric Debye mass squared.
A parametrically larger scale is 
the magnetic one~\cite{linde}, 
$\omega^{ }_\rmiii{M} \sim \alpha \Nc^{ } T $, 
which gives the screening masses 
of spatial correlations in the scalar channel~\cite{ay}.
As such, $\omega^{ }_\rmiii{M}$ determines the magnitude
of finite-volume effects~\cite{finiteV}, but it could 
also affect real-time phenomena. Further frequency scales
originate through $m^{ }_\rmiii{E}$, as well as through 
the thermal ($\omega^{ }_\rmiii{UV} \sim \pi T$) and confinement
($\omega^{ }_\rmi{vac}\sim \Lambdamsbar$) 
scales of the full quantum theory. 

The purpose of the present paper has been to 
carry out numerical simulations within thermally averaged
classical SU(2) and SU(3) lattice gauge theories, in order
to test which of the frequency scales make an appearance 
in the IR part of the hot topological charge density spectral function.  
We have found that this spectral function, or more precisely
the subtracted statistical correlator 
$
 \Delta C^{ }_{\S}(\omega) 
 \stackrel{\rmiii{|\omega| \ll T}}{\approx} 2 T \Delta \rho(\omega) / \omega
$, 
is quadratically growing at 
$\omega \, \sim \, \omega^{ }_\rmiii{IR}$, 
reaching a plateau at 
$\omega \, \sim \, \omega^{ }_\rmiii{M}$ 
(cf.\ \fig\ref{fig:subtr}). 
In terms of \eq\nr{cont_params}, this means that 
\be
 \Delta \rho (\omega) \big|^{ }_\rmi{cont} 
 \;\stackrel{|\omega| \;\ll\; m^{ }_\rmiii{E}}{\simeq}\;
 \kappa\, \omega\, \alpha^5 \Nc^3 (\Nc^2 - 1 ) T^3 \, 
 \frac{1 + 
   \frac{\omega^2}
        { (  c^{ }_\rmiii{IR} \alpha^2 \Nc^2 T )^2_{ } }
 }{ 1 + 
   \frac{\omega^2}{ (  c^{ }_\rmiii{M} \alpha \Nc^{ } T )^2_{ } }
 }
 \;. \la{final_cont}
\ee
Qualitative values for the coefficients 
$\kappa$, $c^{ }_\rmiii{IR}$ and $c^{ }_\rmiii{M}$ can be 
found in \eq\nr{fit_nc3}.
We stress that this representation is of empirical nature, 
as we are not aware of a solid theoretical argument 
for the correct frequency shape 
in the non-perturbative domain. 

Further features are expected to appear 
in $ \rho(\omega) $ and $  C^{ }_{\S}(\omega) $  
if we go to larger frequencies, 
$| \omega | \;\gsim\; m^{ }_\rmiii{E}$. 
However, in this domain no simulations are needed (if $\alpha\Nc^{ } \ll 1$):
the corresponding physics can be addressed 
with a perturbative computation 
in the full quantum theory~\cite{Bulk_wdep}.
For $\omega \sim m^{ }_\rmiii{E}$ this requires 
Hard Thermal Loop resummation, while for $\omega\sim \pi T$ 
a loop computation suffices. 
In our study, this physics 
was addressed by a leading-order perturbative computation within
the classical approximation, whose result was 
subtracted from the lattice measurements. 
The spectral function in the full quantum theory 
can then be estimated by adding together 
the subtracted classical IR 
and full quantum UV parts~\cite{warm}, 
\be
 \rho (\omega) \big|^{ }_\rmi{cont} 
 \; \simeq \;  
 \Delta \rho (\omega) \big|^{ }_\rmi{cont} 
 + 
 \rho^{ }_\rmiii{UV} (\omega) \big|^{ }_\rmi{cont} 
 \;. \la{sum}
\ee 

The IR shape that we have found
can be rephrased by stating that there is no 
transport peak in $\Delta C^{ }_{\S}(\omega) $, but rather 
a ``transport dip'', 
centered at around zero frequency.  
A physical consequence could 
be a rapid thermalization of 
an axion-like inflaton field, as the friction that it feels
is not cut off by the width of a transport peak~\cite{warm}, 
but rather increases if the mass exceeds $\sim \alpha\Nc^{ } T$. 
In the context of non-perturbative estimates of the sphaleron 
rate from imaginary-time lattice simulations~\cite{eucl}, 
such a shape represents possibly a challenge, as a fit to the flat
part of $\Delta C^{ }_{\S}(\omega) $ could lead to an overestimate
of $\Gamma^{ }_\rmi{\rate}$. 

%
\section*{Acknowledgements}

Our work was partly supported 
by the Academy of Finland, 
under grants 320123, 345070 and 319066, 
and 
by the Swiss National Science Foundation
(SNSF), under grant 200020B-188712.
Simulations were carried out at 
CSC, the Finnish IT center for Science, making use
of the newly released HILA suite~\cite{hila}. 

%
\appendix
\renewcommand{\thesection}{\Alph{section}} 
\renewcommand{\thesubsection}{\Alph{section}.\arabic{subsection}}
\renewcommand{\theequation}{\Alph{section}.\arabic{equation}}
%

%
\section{Perturbative evaluation of the spectral function}

The purpose of this appendix is to present 
a perturbative determination of 
$  
 C^\rmi{(cl)}_{\S}(\omega)
$, 
accurate for $\omega \sim 1/a$. 
A convenient method goes through an unlikely detour, 
which however has the advantage that 
we can employ usual perturbative tools, such
as Wick contractions and propagators. 
To this aim, we backtrack to 
the quantized theory; 
turn $\tr[(...)\,e^{-H/T}]$ into a path integral in a spacetime with 
an imaginary-time coordinate $\tau \in (0,\beta)$, 
$\beta \equiv 1/T$; 
represent the Gauss laws 
in \eq\nr{Z_cl}
as integrals over an auxiliary field $\tilde{A}^a_0$; 
and take the classical limit only in the end.\footnote{%
 A benefit of this approach is that as an aside
 we can reproduce the correct leading-order
 quantum-mechanical result, cf.\ footnote~\ref{vac}.
 }  
The spacetime coordinates
are now denoted by $\X \equiv (\tau,\vec{x})$.

With discretized spatial directions, 
the analogue of the continuum field strength,
$
 \tilde F^{a}_{0i} |^{ }_\rmi{cont} \equiv \partial^{ }_0 A^a_i - 
 \mathcal{D}^{ab}_i \tilde{A}^b_0 
$,
can be defined as
\be
 \tilde{F}^{ }_{0i}\, \big|^{ }_\rmi{latt} \; \equiv \; 
 \frac{1}{iag} [\partial^{ }_\tau U^{ }_i(\X)] U^\dagger_i(\X) 
 + 
 \frac{\tilde A^{ }_0(\X) 
 - U^{ }_i(\X) \tilde A^{ }_0(\X+a\vec{i})U^\dagger_i(\X) }{a}
 \;. \la{F0i}
\ee
When we represent the Gauss laws as integrals over 
the Lagrange multiplier $\tilde{A}^a_0$, 
and turn $e^{-H/T}$ into an imaginary-time path integral, the 
dependence of the Euclidean action on $E^{ }_i$, 
incorporating the term from \eq\nr{Ei_1}, takes the form 
\be
 S^{ }_\rmii{E} \; \supset \; 
 \int_0^\beta \! {\rm d}\tau \,
 \sum_\vec{x} a^3 \sum_i \tr
 \bigl[
  (E^{ }_i - i \tilde{F}^{ }_{0i})^2 + \tilde{F}_{0i}^2  
 \bigr]
 \;.
\ee
When we subsequently integrate over the Gaussian fields $E^{ }_i$,
we see that insertions of the operator $E^{ }_i$ get replaced 
by $i\tilde{F}^{ }_{0i}$, 
\be
 \int \! \mathcal{D} E^{ }_i \, (E^{ }_i) \, 
 e^{ - \# (E^{ }_i - i \tilde{F}^{ }_{0i})^2 }
 = 
 \int \! \mathcal{D} E^{ }_i \, 
 (\underbrace{ E^{ }_i - i \tilde{F}^{ }_{0i}}_{\rm odd}
 + i \tilde{F}^{ }_{0i}) \, 
 e^{ - \# (E^{ }_i - i \tilde{F}^{ }_{0i})^2 }
 \;. 
\ee
Therefore, the operator $\chi$ from \eq\nr{chi2} is represented as 
\be
 \chi
 \; \longrightarrow \; 
 \tilde{\chi} 
 \; \equiv \; 
 4 i c^{ }_{\chi} \, \epsilon^{ }_{ijk}\,
 g^2 \tilde{F}^{b}_{0 i} F^{b}_{jk}
 \;. \la{chi_tilde}
\ee 
Less pedantically, 
the same recipe could be obtained from a naive Wick rotation
of $\chi$ from Minkowskian to Euclidean spacetime.


We then 
compute the imaginary-time correlator corresponding to \eq\nr{Delta}, 
\be
 C^{ }_\rmii{E}(\tau) \; \equiv \; 
 \int_\vec{x} \langle \tilde\chi(\tau,\vec{x})\, \tilde\chi(0,\vec{y}) \rangle
 \;, \quad
 0 < \tau < \beta 
 \;. \la{GE}
\ee
After determining the Fourier transform, 
$
 C^{ }_\rmii{E}(\omega^{ }_n) 
 \equiv
 \int_0^{\beta} \! {\rm d}\tau \, 
 e^{i\omega^{ }_n\tau }_{ } C^{ }_\rmii{E}(\tau)
$, 
where 
$\omega^{ }_n = 2 \pi n T$ are Matsubara frequencies, 
we can extract the retarded correlator, 
$
 C^{ }_\rmii{R}(\omega) 
 = 
 C^{ }_\rmii{E}(\omega^{ }_n\to -i [\omega + i 0^+_{ }])
$. 
This in turn permits to determine the statistical correlator
in accordance with \eq\nr{relation},
$
 C^{ }_{\S}(\omega) = [1 + 2 \nB^{ }(\omega)] \im C^{ }_\rmii{R}(\omega)
$.
In order to take the classical limit from \eq\nr{C_cl_w}, 
we recall that had we kept $\hbar$ explicit, 
energy would appear as $\hbar\omega$, so that 
\be
 \nB{ }(\hbar\omega)
 \;
 \stackrel{\hbar\to 0}{\longrightarrow}
 \; 
 \frac{T}{\hbar\omega}
 \;. \la{nB_cl}
\ee
To avoid clutter we suppress $\hbar$ from the notation, 
and simply adopt \eq\nr{nB_cl} at the end of the computation
(cf.\ \eq\nr{cl_lim2}). 

Proceeding with $C^\rmi{ }_\rmii{E}(\omega^{ }_n)$, 
the electric field 
$ \mathcal{\overline{E}}^{\,b}_i $ 
is defined in \eq\nr{E_impr},
and the correspondingly symmetrized  
Euclidean field strength is denoted by
$  \tilde{F}^{b}_{0i} \,|^{ }_\rmi{symm} $. 
We associate $T^b_{ } F^b_{jk} $ with the clover from \eq\nr{Fjk}. 
The links are parametrized as
\be
 U^{ }_j(\X) = e_{ }^{i a g \, T^b_{ } A^b_j (\X)}
 \;, \quad
 A^b_j(\X) = \Tint{P} A^b_j(P)\, e_{ }^{i P\cdot(\X + \frac{a\vec{j}}{2})}
 \;, 
\ee
where 
$\Tinti{P} = T\sum_{p_n}^{} \int_\vec{p}$ 
represents a Matsubara sum-integral, with 
$P = (p^{ }_n,\vec{p})$
and $p^{ }_n = 2\pi n T $; 
and the spatial integral $\int_\vec{p}$ is restricted to a Brillouin zone. 
Adopting the notation 
\be
 \tilde{p}_j^{ } \,\equiv\, \frac{2}{a} \sin \frac{a p_j^{ }}{2}
 \;, \quad
 \ring{p}_j^{ } \,\equiv\, \frac{1}{a} \sin {a p_j^{ }}
 \;, \quad
 \undertilde{p_j} \,\equiv\, \cos \frac{a p_j^{ }}{2}
 \;, \quad
 \ring{p}^{ }_j = \tilde{p}^{ }_j \undertilde{p_j}
 \;, \la{momenta}
\ee
the electric and magnetic fields can be expressed as 
\ba
 \tilde{F}^{b}_{0i}(\X) \, \big|^{ }_\rmi{symm} 
 & = &  
 i \, \Tint{P} 
 e_{ }^{i P\cdot  \X }
 \, 
 \undertilde{p^{ }_i}
 \bigl[\, 
   p^{ }_n A^{b}_i(P) 
 - \tilde{p}^{ }_i \tilde{A}^{b}_0(P)
 \,\bigr]
 \; + \; 
 \rmO(g)
 \;, \\
 {F}^{b}_{jk}(\X) 
 & = &  
 i\, \Tint{P} 
 e_{ }^{i P\cdot \X}
 \, 
 \bigl[ 
   \ring{p}^{ }_j \undertilde{p^{ }_k} A^{b}_k(P) 
 - 
   \ring{p}^{ }_k \undertilde{p^{ }_j} A^{b}_j(P) 
 \bigr]
 \; + \; 
 \rmO(g)
 \;.
\ea
Then the topological charge density from \eq\nr{chi_tilde} becomes 
\be
 \tilde{\chi}(\X)   =   
 8 i c^{ }_\chi \epsilon^{ }_{ijk}\, g^2 \, 
 \Tint{P,Q} \!\! 
 e_{ }^{i (P+Q) \cdot  \X }
 \undertilde{p^{ }_i}
 \bigl( \delta^{ }_{\mu i}\, p^{ }_n
      - \delta^{ }_{\mu 0}\, \tilde{p}^{ }_i \bigr)
 \, \ring{q}^{ }_k \undertilde{q^{ }_j}
 A^b_\mu(P) A^b_j(Q) 
 \; + \; 
 \rmO(g^3)
 \;, 
\ee
where Greek indices take values
$\mu\in\{0,1,2,3\}$; 
Latin indices are spatial,  
$i,j,k\in\{1,2,3\}$; 
and repeated indices are summed over.

Carrying out the contractions for \eq\nr{GE}, 
the Fourier transform takes the form
\ba
 C^{ }_\rmii{E}(\omega^{ }_n)
 & = & 
 - 64 \dA^{ }c_\chi^2\, g^4_{ } T\sum_{p^{ }_n,q^{ }_n} 
   \delta^{ }_{0,\,\omega^{ }_n + p^{ }_n + q^{ }_n}
   \int_\vec{p} 
   \epsilon^{ }_{ijk} \epsilon^{ }_{suv}
   \undertilde{p^{ }_i}
   \undertilde{p^{ }_j}\ring{p}^{ }_k
   \undertilde{p^{ }_s}
   \undertilde{p^{ }_u}\ring{p}^{ }_v 
   \bigl(   \delta^{ }_{\mu i}\, p^{ }_n
          - \delta^{ }_{\mu 0}\,\tilde{p}^{ }_i \bigr)
 \nn 
 & \times &
 \Bigl\{ 
   \Delta^{ }_{\mu \alpha}(p^{ }_n,\vec{p})\, 
   \Delta^{ }_{j u}(q^{ }_n,-\vec{p})
   \, 
   \bigl(
     \delta^{ }_{\alpha s}\, p^{ }_n - \delta^{ }_{\alpha 0}\, \tilde{p}^{ }_s 
   \bigr)
 \nn 
 & & \; - \, 
   \Delta^{ }_{\mu u}(p^{ }_n,\vec{p})\, 
   \Delta^{ }_{j \alpha}(q^{ }_n,-\vec{p})
   \, 
   \bigl(
    \delta^{ }_{\alpha s}\, q^{ }_n + \delta^{ }_{\alpha 0}\, \tilde{p}^{ }_s 
   \bigr)
 \Bigr\}  
 + \rmO(g^6)
 \;, 
\ea
where $\dA^{ }\equiv \Nc^2 - 1$, and   
$\Delta^{ }_{\mu\nu}$ is the gauge field propagator.

As a first check, it can be verified that longitudinal parts of the 
propagators, 
$
 \Delta^{ }_{\alpha\beta}(P) \supset
 c\, \tilde{p}^{ }_\alpha\tilde{p}^{ }_\beta
$, 
where $\tilde{p}^{ }_0 \equiv p^{ }_n$, 
yield no contributions. This is fairly simple in the three 
cases where a ``projector'', e.g.\  
$
   \delta^{ }_{\alpha s}\, p^{ }_n - \delta^{ }_{\alpha 0}\, \tilde{p}^{ }_s 
$, 
is contracted with a propagator, e.g.\ 
$
 \Delta^{ }_{\mu\alpha}(P)
$.
More work, making use of the antisymmetry of 
the Levi-Civita symbols, 
is required for the fourth propagator, 
{\it viz.}\ 
$
 \Delta^{ }_{j u}(q^{ }_n,-\vec{p})
$, 
to see that its longitudinal part does not contribute either. 

Given that the longitudinal parts do not contribute, propagators
can be replaced by their Feynman parts, 
$
 \Delta^{ }_{\alpha\beta}(P) \to 
 \delta^{ }_{\alpha\beta} / (p_n^2 + \tilde{p}^2 )
$.
Here we have introduced the notation 
\be
 \tilde{p}^{2n} \; \equiv \; \sum_{j=1}^3 \tilde{p}_j^{2n}
 \;, \quad
 n = 1,2, ... 
 \;. \la{powers}
\ee
We then end up with  
\ba
 C^{ }_\rmii{E}(\omega^{ }_n)
 & = & 
 - 64 \dA^{ }c_\chi^2\, g^4_{ } T\sum_{p^{ }_n,q^{ }_n} \int_\vec{p}
   \delta^{ }_{0,\,\omega^{ }_n + p^{ }_n + q^{ }_n}
   \frac{ 
  2  (p^{2}_n + p^{ }_n q^{ }_n) \, \tilde{p}^2_{ } \, 
  \Theta(\vec{p})
   }{
   (p_n^2 + \tilde{p}^2)(q_n^2 + \tilde{p}^2) 
   }
 \;, \la{CE1} \\ 
 \Theta(\vec{p}) & = &
   \undertilde{p^{2}_1}
   \undertilde{p^{2}_2}
   \undertilde{p^{2}_3}
    \; = \; 
  \biggl( 1 - \frac{a^2\tilde{p}_1^2}{4} \biggr)
  \biggl( 1 - \frac{a^2\tilde{p}_2^2}{4} \biggr)
  \biggl( 1 - \frac{a^2\tilde{p}_3^2}{4} \biggr)
  \nn  
  & = &  
  1  - \frac{a^2 \tilde{p}^2_{ }}{4}
  + \frac{a^4 [(\tilde{p}^2_{ })^2_{ } - \tilde{p}^4_{ }]}{32}
  - \frac{a^6  
  [( \tilde{p}^2_{ } )^3_{ }
  - 3 \tilde{p}^2_{ }\tilde{p}^4_{ }
   + 2 \tilde{p}^6_{ }
  ]}{384}
 \;.
 \la{Theta}  
\ea
Furthermore, by symmetry, we may replace
$
 2 (p_n^2 + p^{ }_n q^{ }_n) \to (p^{ }_n + q^{ }_n)^2 = \omega^2_n
$.

Subsequently, writing
$
 \delta^{ }_{0,\,\omega^{ }_n + p^{ }_n + q^{ }_n} 
 = T \int_0^{\beta}\! {\rm d}\tau \,
   e^{i (\omega^{ }_n + p^{ }_n + q^{ }_n)\tau}
$, 
the two Matsubara sums 
in \eq\nr{CE1} 
can be carried out,  
\be
 T \sum_{p^{ }_n} 
 \,\frac{
  e^{i p^{ }_n \tau}
 }{p_n^2 + \tilde{p}^2}
 = \frac{\nB^{ }(\tilde{p})}{2\tilde{p}}
 \Bigl[
  e^{\tilde{p}\tau}_{ } + 
  e^{\tilde{p}(\beta-\tau)}_{ } 
 \Bigr]
 \;, \la{sums}
\ee
where $\tilde{p} \equiv \sqrt{\tilde{p}^2}$. 
When we multiply structures like in \eq\nr{sums} 
with each other, cross terms are independent of $\tau$, 
and yield no contribution in the end. 
For one of the remaining terms, 
the Fourier transform yields
\be
 \int_0^{\beta} \! {\rm d}\tau \, e^{i\omega^{ }_n\tau} 
 \, \nB^2(\tilde{p})\, e_{ }^{2(\beta - \tau)\tilde{p}}
 \; = \; 
 \frac{\nB^2(\tilde{p})(1 - e^{2\beta\tilde{p}}_{ })}
 {i\omega_n^{ }- 2 \tilde{p}}
 \; = \; 
 - \, \frac{1 + 2 \nB^{ }(\tilde{p})}{i\omega_n^{ }- 2 \tilde{p}}
 \;.
\ee
The cut then gives 
\be
 \im \biggl\{ \frac{1}{i\omega_n^{ }- 2 \tilde{p}}
 \biggr|^{ }_{\omega^{ }_n\to -i [\omega + i 0^+_{ }]}
 \biggr\}
 \; = \; 
 - \pi \delta(\omega - 2 \tilde{p})
 \;. 
\ee
All in all, this implies that 
\ba
 T\sum_{p^{ }_n,q^{ }_n} 
   \frac{ 
   \delta^{ }_{0,\,\omega^{ }_n + p^{ }_n + q^{ }_n}
   }{
   (p_n^2 + \tilde{p}^2)(q_n^2 + \tilde{p}^2) 
   }
 & \stackrel{\rmii{cut}}{\longrightarrow} & 
  \frac{
  \pi [ \delta(\omega - 2 \tilde{p}) + \delta(\omega + 2 \tilde{p}) ]
      [ 1 + 2 \nB^{ }(\tilde{p}) ] 
  }{4\tilde{p}^2}
  \;. \la{cut}
\ea

The final step is to consider 
the classical limit from \eq\nr{C_cl_w}.\footnote{%
 If we take no classical limit but go to continuum, 
 where $\tilde{p}\to p$ and $a\to 0$, then 
 \eqs\nr{CE1}--\nr{cut} reproduce
 an expression employed in ref.~\cite{warm}, 
 \be
  \im C^{ }_\rmiii{E} (\omega^{ }_n\to -i [\omega + i 0^+_{ }])
  = 64 \dA^{ }c_\chi^2\, g^4_{ } 
  \int_\vec{p} p^2 [1 + 2 \nB^{ }(p)] \pi \delta(\omega - 2p) 
  = 
  \frac{\dA^{ }c_\chi^2\, g^4_{ } \omega^4}{\pi}
  \Bigl[1 + 2\, \nB^{ }\Bigl( \frac{\omega}{2} \Bigr) \Bigr] 
  \;. \nonumber
 \ee
 \la{vac}
 }  
Two Bose distributions appear, one from 
$
 C^{ }_{\S}(\omega) = [1 + 2 \nB^{ }(\omega)]
 \,\rho 
 (\omega)
$, 
cf.\ \eq\nr{relation}, 
and the other from \eq\nr{cut}.
Reinstituting~$\hbar$, we are faced with
\be
 \lim_{\hbar \to 0}  
 \bigl[ 1 + 2 \nB^{ }(\hbar\omega) \bigr]
 \Bigl[ 1 + 2 \nB^{ }\Bigl( \frac{\hbar \omega}{2} \Bigr)
 \Bigr] = \frac{2 T}{\hbar\omega} \frac{4 T}{\hbar\omega} 
 \;. \la{cl_lim2}
\ee
Then, from \eqs\nr{CE1}, \nr{Theta} and \nr{cut}, 
\ba
  C^\rmi{(cl)}_{\S}(\omega)
 & = &  
  64 \dA^{ }c_\chi^2\, g^4 T^2 
  \int_{\vec{p}} \pi \delta\Bigl(\tilde{p} - \frac{\omega}{2} \Bigr)
  \Theta(\vec{p})
 \;. \la{Ccl1}
\ea

Proceeding to the numerical evaluation,
we employ dimensionless units, writing 
\be
 p^{ }_i \equiv \frac{q^{ }_i}{a}
 \;, \quad
 \tilde{p}^{ }_i \equiv \frac{\tilde{q}^{ }_i}{a}
 \;. 
\ee
Normalizing like in \eq\nr{normalization}, 
\eq\nr{Ccl1} becomes 
\be
 \frac{ C^\rmi{(cl)}_{\S}(\omega) }{(\alpha T)^4}
 \; \stackrel{a g^2 T \ll a\omega }{\approx} \;
 \frac{4 \dA^{ }}{ (a g^2 T)^2 } 
 \,
 \mathcal{C}\Bigl( \frac{a\omega}{2} \Bigr) 
 \;, \la{a4Ccl3}
\ee
where the frequency range is the one in which corrections
from higher loop orders are small. The function 
$\mathcal{C}$ has been defined as 
\ba
 \mathcal{C}( x ) 
 & \equiv &  
 \int_0^{\pi} \! \frac{{\rm d}^3\vec{q}}{\pi^2}
 \, \delta\Bigl( \tilde{q} - x \Bigr) 
 \, \Theta\Bigl( \frac{\vec{q}}{a} \Bigr)
 \nn 
 & = & 
 \frac{2}{\pi^2}
 \int_0^\pi \! {\rm d}q^{ }_1 \! 
 \int_0^{q^{ }_1} \! {\rm d}q^{ }_2 \! 
 \int_0^{\pi} \! {\rm d}q^{ }_3 \, 
 \delta( \tilde{q} - x )
 \, \Theta\Bigl( \frac{q^{ }_1}{a},
                 \frac{q^{ }_2}{a},
                 \frac{q^{ }_3}{a} \Bigr) 
 \;. 
\ea
Here we made use of the symmetry of $\Theta$
in $p^{ }_1 \leftrightarrow p^{ }_2$.
The last step is 
to carry out the integral 
over $q^{ }_3$, which yields
[here $\phi$ is an arbitrary function, 
$\phi(q^{ }_3) = \Theta ( \frac{q^{ }_1}{a},
                 \frac{q^{ }_2}{a},
                 \frac{q^{ }_3}{a} )$]
\ba
 \int_0^{\pi} 
 \! {\rm d}q^{ }_3 \, \delta( \tilde{q} - x ) \, \phi(q^{ }_3)
 & = & 
 \theta \biggl(
 \frac{ x^2 }{4} 
 - \sin^2 \frac{q^{ }_1}{2} 
 - \sin^2 \frac{q^{ }_2}{2} 
 \biggr)
 \theta \biggl(
 1 
 + \sin^2 \frac{q^{ }_1}{2} 
 + \sin^2 \frac{q^{ }_2}{2} 
 - \frac{ x^2 }{4} 
 \biggr)
 \nn 
 & \times & 
 \frac{ x }{|\sin(q^{ }_3)|}
 \, 
 \phi\biggl( 
 2 \arcsin\sqrt{
 \frac{ x^2 }{4} 
 - \sin^2 \frac{q^{ }_1}{2} 
 - \sin^2 \frac{q^{ }_2}{2} 
 } \; 
 \biggr) 
 \;, 
\ea
where $\theta$ is the Heaviside function.\footnote{%
 The integration boundaries for $q^{ }_1$ and $q^{ }_2$
 are modified by the two Heaviside functions. 
 } 
The argument of $\phi$ permits for us to evaluate 
$
  {x} / {|\sin(q^{ }_3)|}
$
as well as powers like $\tilde{q}^{2n}_{ }$.
Numerical illustrations 
can be found as curves in \fig\ref{fig:bare}.

\small{
%

}

\end{document}